\documentclass[11pt]{article}
\usepackage[final]{acl}
\usepackage{times,latexsym}
\usepackage[T1]{fontenc}
\usepackage[utf8]{inputenc}
\usepackage{microtype}
\usepackage{amsmath,amssymb,graphicx}
\usepackage{algorithm}
\usepackage{algpseudocode}
\usepackage{tabularx}
\usepackage{enumitem}
\usepackage{capt-of}

\usepackage{booktabs,multirow,array}
\usepackage{placeins}
\usepackage[table]{xcolor}
\definecolor{tblHeader}{HTML}{DCE8F3}
\definecolor{tblBaseline}{HTML}{F5F6F7}
\definecolor{tblCore}{HTML}{EDF4FA}
\definecolor{tblHalting}{HTML}{EAF5EE}
\definecolor{tblOracle}{HTML}{ECEFF1}
\definecolor{tblRegress}{HTML}{FBECEA}
\definecolor{tblAccent}{HTML}{2F5F89}
\definecolor{tblGreen}{HTML}{347354}
\definecolor{tblRule}{HTML}{000000}

\newcommand{\tablebaselinerow}{\rowcolor{tblBaseline}}
\newcommand{\tablemethodrow}{\rowcolor{tblCore}}
\newcommand{\tablebestrow}{\rowcolor{tblHalting}}
\newcommand{\tableregressrow}{\rowcolor{tblRegress}}
\arrayrulecolor{tblRule}

\newcommand{\method}{CoReLoop}

\newcommand{\h}{\mathbf{h}}
\newcommand{\s}{\mathbf{s}}
\newcommand{\e}{\mathbf{e}}
\newcommand{\g}{\mathbf{g}}

\newcommand{\cstate}{\widetilde{\mathbf{s}}}
\newcommand{\Aset}{\mathcal{A}}
\newcommand{\Lcal}{\mathcal{L}}

\title{CoReLoop: Parameter-Efficient Controlled Recurrent Refinement for Audio Deepfake Detection}

\author{
Kunyu Feng$^{1}$,
Yuxiang Wang$^{1}$,
Li Wang$^{1}$,
Wan Lin$^{1}$,
Zhizheng Wu$^{1,2,*}$ \\
$^{1}$The Chinese University of Hong Kong, Shenzhen \\
$^{2}$Amphion Technology Co., Ltd. \\
{\small
\texttt{kunyufeng@link.cuhk.edu.cn} \quad
\texttt{wuzhizheng@cuhk.edu.cn}
}
{\small $^{*}$Corresponding author.}
}

\begin{document}
\maketitle

\begin{abstract}
Audio deepfake detectors are expected to remain robust to diverse attacks, datasets,
and recording conditions, yet collecting training data that cover every
future condition is impractical. We investigate whether an already-trained
SSL-based detector can improve its generalization through additional
recurrent computation, without introducing new training data or updating
its original parameters. Directly feeding the encoder output back as its
next input, however, severely degrades detection, revealing a mismatch
between the representations produced and consumed by a model trained for
single-pass inference. We therefore propose \method{}, which enables
controlled recurrent refinement through loop-specific LoRA and lightweight
modules that adapt recurrent inputs, regulate state updates, and align
refined representations with the frozen classifier. Across 14 diverse test
sets, \method{} improves pooled EER across 3-, 6-, 12-, and 24-layer
encoders; on the 24-layer model, two-pass refinement reduces EER from
4.85\% to 3.74\% while training only approximately 11M of 598M parameters.
An optional halting head further selects refinement depth per utterance,
achieving 3.73\% EER with an average of 1.18 passes. Analysis shows that later-pass saturation coincides with smaller
applied updates and nearly balanced corrections and regressions,
while the recurrent state continues to change.
\end{abstract}

\begin{figure}[t]
    \centering
    \includegraphics[width=0.95\columnwidth]{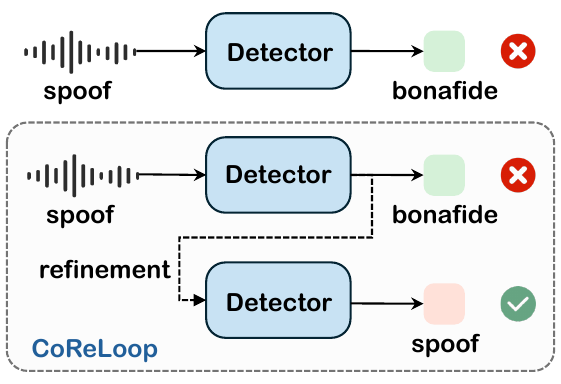}
    \caption{
    Motivation for recurrent refinement.
A spoofed utterance misclassified after a single pass can be corrected
through additional recurrent processing.
    }
    \label{fig:intro_motivation}
\end{figure}

\section{Introduction}
\label{sec:intro}

Audio deepfake detection aims to distinguish genuine speech from speech
generated by synthesis or voice conversion. A common detector combines a
self-supervised learning (SSL) speech encoder with a classifier that predicts
authenticity from the encoded representation~\citep{tak22_odyssey}.
These detectors are expected to remain robust to novel utterances,
generators, recording conditions, and datasets whose acoustic distributions
may differ from the training data~\citep{muller24b_interspeech}.
As synthesis and voice conversion methods continue to evolve, collecting
training data that cover every future condition is
impractical~\citep{zhang2021one}. This motivates seeking better
generalization without relying on broader attack coverage during training.
We investigate whether an already-trained detector can improve upon its
single-pass performance under diverse test conditions through
additional refinement,
without expanding the training data or updating its original parameters.

To enable additional refinement, we consider recurrence, where the
representation produced by one encoder pass guides the next. This feedback
allows subsequent processing to build on previously extracted features,
while parameter sharing increases effective depth without duplicating the
encoder. In our setting, however, the encoder has already been optimized for
single-pass inference, and its original parameters remain frozen.
Introducing recurrence therefore requires converting this existing
feed-forward computation into an effective iterative refinement process.

This conversion requires reconciling the representations produced and
consumed by the encoder. Its first layer was trained to receive acoustic
frontend features, whereas direct recurrence supplies the final layer's
output in their place. In our diagnostic, simply feeding this output back
without adaptation raises pooled EER from 4.85\% at the first pass to
10.52\% and 64.68\% at the next two. These results show that additional
encoder evaluations alone do not guarantee better detection. With the
original detector frozen, effective recurrence therefore calls for
mechanisms that construct suitable recurrent inputs, control state updates,
and maintain compatibility with the classifier.

We propose \method{} (Controlled Refinement Loop), which enables recurrent
refinement in a frozen detector through loop-specific LoRA and lightweight
state-control modules. Only the added components are trained on the
original data, while the original detector remains frozen. All refinement
modules, including LoRA, are bypassed at $T=1$ to preserve the detector's
original single-pass behavior, retaining an unchanged prediction path
when refinement is disabled. An optional halting head determines which
utterances receive additional processing and selects their refinement
depth during inference.

Across 14 diverse test sets, \method{} improves pooled EER for
3-, 6-, 12-, and 24-layer encoders. On the 24-layer model, two-pass
refinement reduces pooled EER from 4.85\% to 3.74\% while training
about 11M of 598M parameters. Adaptive inference achieves 3.73\% EER
with 1.18 passes on average, using 41\% fewer core evaluations than
fixed two-pass inference. Most gains come from the first additional
pass; later saturation accompanies attenuated updates and nearly
balanced corrections and regressions. Learned halting preserves more
recoverable errors than random or confidence-based routing, while a
training--test overlap audit clarifies the scope of the generalization gains.
Our contributions are:
\begin{itemize}[leftmargin=*,noitemsep,topsep=2pt]
    \item We introduce \method{}, a parameter-efficient framework that
    retrofits recurrent refinement into a frozen audio deepfake detector
    through recurrent-input alignment, state control, and output alignment.

    \item We demonstrate that recurrent refinement improves detection
    across diverse evaluation conditions and encoder depths, while adaptive
    halting provides sample-dependent computation without modifying the
    original detector.

    \item We analyze recurrent refinement through component ablations,
    decision transitions, hidden-state tracing, and deeper unrolling
    to characterize the observed later-pass saturation.

    \item We analyze adaptive routing and training--test overlap to
    understand where additional computation is useful and to clarify the
    scope of the resulting generalization claims.
\end{itemize}

\section{Related Work}
\label{sec:related}

\subsection{Generalizable audio deepfake detection}

Audio deepfake detection has progressed from waveform- and
spectrogram-based architectures, such as RawNet2 and
AASIST~\citep{tak2021endtoend,jung2022aasist}, to systems combining
pretrained SSL speech encoders with lightweight
classifiers~\citep{tak22_odyssey}. SSL representations provide strong
acoustic features, but detectors trained on known attacks often degrade
under unseen synthesis methods, recording conditions, languages, and
transmission channels~\citep{muller24b_interspeech,11345101}. Existing
methods commonly improve generalization through waveform augmentation,
multi-source training, or representations designed to reduce dependence
on dataset-specific artifacts~\citep{tak2022rawboost}. These approaches
primarily strengthen the detector by broadening or modifying its training
distribution. In contrast, \method{} investigates whether an
already-trained detector can improve through recurrent refinement,
without additional attack data or updates to its original parameters.

\begin{figure*}[t]
\centering
\includegraphics[width=\textwidth]{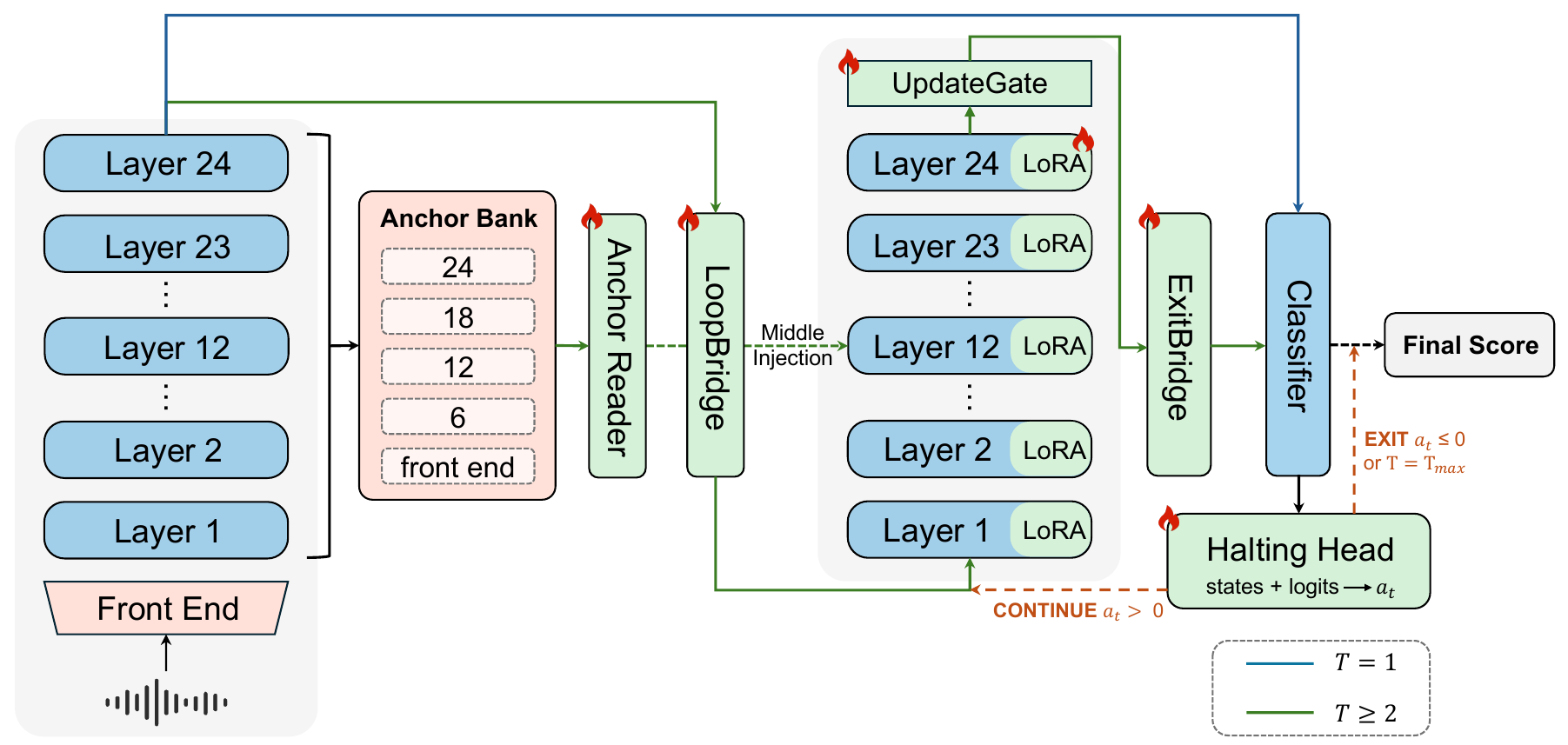}
\caption{Overview of \method{}. At $T=1$, loop-specific modules are bypassed,
yielding the frozen baseline path. For $T\geq2$, lightweight adaptation and
state-control modules refine the recurrent state, while the halting head
determines whether each sample should exit or continue.}
\label{fig:overview}
\end{figure*}

\subsection{Recurrent-depth refinement}

Recurrent-depth models repeatedly apply shared layers to increase effective
depth without proportional parameter growth. Universal Transformers and
recent recurrent-depth language models demonstrate that shared computation
can support iterative representation refinement when explicitly trained
for repeated processing~\citep{dehghani2019universal,
geiping2025scalingtesttimecomputelatent,wang2026recurtrace}. Additional
iterations can therefore improve prediction quality while reusing the same
parameters. Our setting is different: recurrence is retrofitted into a
frozen speech encoder originally optimized for a single forward pass.
Naively repeating its internal layers may produce states that are
incompatible with the expected encoder interfaces, motivating explicit
input and output alignment during refinement. REIMU applies recurrence
within a downstream network over fixed SSL features~\citep{
ng2026reimuefficientheterogeneoushierarchical}, whereas \method{} reuses
layers inside the existing SSL encoder and preserves its original
first-pass prediction as the baseline path.

\subsection{Parameter-efficient adaptation and adaptive computation}

LoRA enables efficient adaptation of pretrained models by introducing
low-rank updates while keeping the original parameters
frozen~\citep{hu2022lora}. It is commonly used to adapt a model to a new
task or domain. In \method{}, LoRA instead specializes shared encoder
layers for recurrent refinement, while the original single-pass inference
path remains unchanged. This separation allows additional computation to
modify recurrent behavior without perturbing the frozen baseline.
Adaptive-computation methods, including Adaptive Computation Time and
PonderNet, allocate different numbers of processing steps according to
input difficulty~\citep{graves2016adaptive,banino2021pondernet}. Following
this principle, our optional halting head determines after each pass
whether an utterance should exit or receive further refinement, enabling
sample-dependent computation without updating model parameters during
inference.

\section{Method}
\label{sec:method}

\subsection{Frozen detector and recurrent inputs}
\label{ssec:formulation}

A trained detector comprises an acoustic frontend $P_{\theta_p}$, an encoder
core $F_{\theta_f}$ containing all retained layers, and a pooling--classification
module $C_{\theta_c}$. All original parameters remain frozen. The first pass
produces frontend features $\h_0$ and encoder output $\s_1$, which the frozen
classifier processes directly. All refinement modules are bypassed during
the first pass, thereby exactly preserving the original detector's prediction.

For $t\geq2$, a LoopBridge maps the previous endpoint $\s_{t-1}$ back
into the core input space, using $\h_0$ as a fixed Prelude anchor.
Conditioned on first-pass anchor context $\e_t$ and a step embedding
$\tau_t$, it constructs the recurrent input as
\begin{equation}
\begin{aligned}
  \h_t &= \h_0
  + B_{\Delta}\!\left(\operatorname{Norm}(\s_{t-1}-\h_0)\right)\\
  &\quad + B_e(\e_t) + B_{\tau}(\tau_t),
\end{aligned}
\label{eq:loop_bridge}
\end{equation}
where $\operatorname{Norm}$ denotes normalization, and
$B_{\Delta}$, $B_e$, and $B_{\tau}$ are lightweight trainable
projections into the core input space. This construction conditions
the recurrent input on the previous core state's deviation from
the Prelude anchor.

To provide context $\e_t$, we construct an anchor bank $\Aset$
from first-pass representations using source-specific LayerNorm
and linear projections from 1024 to 256 dimensions. The 24-layer
model uses the frontend output and the hidden states at layers
$\{6,12,18,24\}$, yielding five anchor sources. The bank is
bypassed at $t=1$ and remains fixed throughout refinement. For $t\geq2$, an
AnchorReader conditioned on $\s_{t-1}$ and $\tau_t$ reads across
anchor layers and frames to obtain $\e_t$. This context is also
injected after layer 12 in the 24-layer model.

\subsection{Controlled recurrent refinement}
\label{ssec:controlled_update}

Later passes activate loop-specific LoRA updates $\Delta\theta_t$ on
attention projections (q/k/v/o) and FFN linear layers. Each adapted
frozen linear map receives a rank-$r$ update with trainable factors
$A$ and $B$ shared across passes, scaled by $\alpha/r$ and a learned
step-specific multiplier $\gamma_t$. The shared encoder produces a candidate state, which is then blended with
the previous endpoint through a learned update gate at each refinement pass:
\begin{equation}
\begin{aligned}
  \cstate_t &= F_{\theta_f,\Delta\theta_t}(\h_t;\e_t),\\
  \s_t &= \s_{t-1}+\g_t\odot(\cstate_t-\s_{t-1}).
\end{aligned}
\label{eq:update}
\end{equation}
The UpdateGate predicts an utterance-conditioned, channel-wise
sigmoid gate $\mathbf{g}_t\in(0,1)^H$ from mask-aware temporal
summaries of the previous and candidate states, their difference,
the anchor context, and the step embedding. Each channel has its
own gate value, shared across all frames of the utterance. For any terminal depth $t>1$, an ExitBridge applies a lightweight
residual projection to $\s_t$, conditioned on $\s_1$ and $\tau_t$,
to align the refined state with the frozen classifier. For concise
notation, we define $\bar{\s}_1=\s_1$ and
$\bar{\s}_t=E_{\Omega}^{\mathrm{exit}}(\s_t;\s_1,\tau_t)$ for $t\geq2$,
where $\Omega$ denotes all trainable refinement parameters. The
depth-$t$ classification logits are then
$\mathbf{z}_t=C_{\theta_c}(\bar{\s}_t)$.

\begin{algorithm}[t]
\caption{Two-stage training of \method{}.}
\label{alg:training}
\small
\begin{algorithmic}[1]
\Require Training set $\mathcal D$ and frozen detector

\Statex \textbf{Stage I: Recurrent refinement}
\For{each minibatch $(\mathbf{x},y)\sim\mathcal D$}
    \State Run the frozen first pass and construct the anchor bank
    \State Sample $T\in\{2,3\}$ and unroll refinement to depth $T$
    using Eqs.~\eqref{eq:loop_bridge}--\eqref{eq:update}
    \State Update refinement parameters using terminal BCE
\EndFor

\Statex \textbf{Stage II: Adaptive halting}
\State Freeze the recurrent refiner
\For{each minibatch $(\mathbf{x},y)\sim\mathcal D$}
    \State Unroll to depths $1$--$3$ without gradients
    \State Construct continuation targets using
    Eq.~\eqref{eq:continue_targets}
    \State Update the halting head using
    Eq.~\eqref{eq:halting_loss}
\EndFor
\end{algorithmic}
\end{algorithm}

\subsection{Training and adaptive halting}
\label{ssec:training_halting}

Training proceeds in two stages, as summarized in
Algorithm~\ref{alg:training}. Let $\Omega$ collect the parameters of
the loop-specific LoRA, LoopBridge, anchor-bank projections,
AnchorReader, UpdateGate, ExitBridge, and step embeddings. In Stage I,
only $\Omega$ is optimized on the original training data, while all
detector parameters remain frozen. Each batch uniformly samples
$T\in\{2,3\}$ and minimizes BCE on the terminal prediction
$\mathbf{z}_T$.

In Stage II, we freeze both the original detector and $\Omega$, and
train only the halting head $H_\psi$. 
Evaluating each sample at all three depths without gradients gives its
per-depth BCE losses $\ell_t$, computed from the classification logits.
At $t\in\{1,2\}$, $H_\psi$ predicts a continue logit $a_t$ from the current and
previous endpoints and logits; at $t=1$, $\h_0$ and a zero logit provide the
previous inputs. Continuation is supervised when a later pass reduces the
loss by more than $\delta=0.01$:
\begin{equation}
  c_t=\mathbb{I}\!\left[
    \min_{t<k\leq3}\ell_k+\delta<\ell_t
  \right],\qquad t\in\{1,2\}.
\label{eq:continue_targets}
\end{equation}

With $p_t=\sigma(a_t)$, the soft exit probabilities are
$q_1=1-p_1$, $q_2=p_1(1-p_2)$, and $q_3=p_1p_2$,
which define a differentiable halting distribution. We minimize
\begin{equation}
\begin{aligned}
  \Lcal_{\mathrm{halt}}
  &=\sum_{t=1}^{2}\mathrm{BCE}_{w_t^+}(a_t,c_t)\\
  &\quad+0.5\,\mathbb{E}\!\left[\sum_{t=1}^{3}q_t\ell_t\right].
\end{aligned}
\label{eq:halting_loss}
\end{equation}
Here, $w_t^+$ is an inverse-frequency positive-class weight capped at $20$.
The second term minimizes classification loss under the induced exit
distribution, helping mitigate the mismatch between all-depth supervision
during training and sequential halting at inference.
At inference, $a_t>0$ continues refinement and $a_t\leq0$ exits; the third
pass always exits in the main evaluation, giving $T_{\max}=3$. The separate
depth-extrapolation diagnostic in Table~\ref{tab:depth6} unrolls the same
refiner to six passes and applies the $T_{\max}=6$ halting configuration.

\section{Experiments}
\label{sec:experiments}

\begin{table*}[t]
\centering
\caption{EER (\%) across 14 diverse test sets and RTF on 24-layer
W2V-BERT (seed 42; 598M parameters, 11M trainable). Pooled EER
(primary metric) uses a pooled threshold; individual EERs use
test-set-specific thresholds. Macro averages EER over 14 test sets.
RTF is inference time divided by audio duration. Lower is better;
best non-oracle results are bolded. Random, Confidence, Halting,
and Oracle are \method{} policies; Confidence uses a threshold of $0.9$,
while Oracle is a non-deployable ideal halting policy.}
\label{tab:main}
\begingroup
\setlength{\tabcolsep}{3pt}
\renewcommand{\arraystretch}{1.16}
\arrayrulecolor{tblRule}
\def\coreLoopMainTableBody{%
\toprule
\multirow{2}{*}{\textbf{Model}}
& \textbf{RTF}
& \multirow{2}{*}{\textbf{Pooled}}
& \multirow{2}{*}{\textbf{Macro}}
& \multirow{2}{*}{\textbf{ITW}}
& \multicolumn{4}{c}{\textbf{ASVspoof}}
& \multirow{2}{*}{\textbf{FoR}}
& \multirow{2}{*}{\textbf{CF}}
& \multicolumn{4}{c}{\textbf{ADD}}
& \multirow{2}{*}{\textbf{DFADD}}
& \multirow{2}{*}{\textbf{LSV}}
& \multirow{2}{*}{\textbf{SONAR}}\\
\cmidrule(lr){6-9}
\cmidrule(lr){12-15}
& $10^{-2}$
& & &
& \textbf{19}
& \textbf{21LA}
& \textbf{21DF}
& \textbf{24}
& &
& \textbf{22T1}
& \textbf{22T3}
& \textbf{23R1}
& \textbf{23R2}
& & &\\
\midrule

\tablebaselinerow
W2V-BERT ($T=1$)
& \textbf{1.71}
& 4.85 & 2.86 & 1.88 & \textbf{0.34} & 2.73 & 2.69 & 10.43
& 2.15 & 0.22 & 12.45 & 0.90 & 1.63 & 4.51
& \textbf{0.00} & \textbf{0.13} & \textbf{0.00}\\

LoRA ($T=1$)
& 2.44
& 4.24 & 2.81 & 2.20 & 0.87 & 3.27 & 2.15 & \textbf{9.28}
& 2.15 & 0.45 & 10.88 & 0.89 & 1.31 & \textbf{3.99}
& \textbf{0.00} & 0.63 & 1.20\\

\midrule

\tablemethodrow
{\method{}} ($T=2$)
& 3.63
& 3.74 & \textbf{2.47} & 1.66 & 0.63 & \textbf{2.26}
& 2.18 & 9.53 & \textbf{0.86} & \textbf{0.21}
& \textbf{10.55} & \textbf{0.81} & 1.36 & 4.20
& \textbf{0.00} & 0.38 & \textbf{0.00}\\

\tablemethodrow
{\method{}} ($T=3$)
& 5.47
& \textbf{3.69} & 2.49 & 1.24 & 0.62 & 2.39
& 2.18 & 9.56 & 1.28 & 0.22
& 10.60 & 0.92 & \textbf{1.25} & 4.20
& \textbf{0.00} & 0.44 & \textbf{0.00}\\

{Random} ($\overline{T}=1.75$)
& 3.31
& 4.03 & 2.75 & 2.05 & 0.77 & 2.72
& 2.04 & 9.72 & 1.73 & 0.35
& 11.53 & 1.34 & 1.69 & 4.14
& \textbf{0.00} & 0.44 & \textbf{0.00}\\

{Confidence} ($\overline{T}=1.06$)
& 1.77
& 4.08 & 2.71 & 1.75 & 0.52 & 2.27
& \textbf{2.03} & 9.86 & 1.73 & 0.52
& 11.94 & 0.87 & 1.49 & 4.23
& \textbf{0.00} & 0.69 & \textbf{0.00}\\

\tablebestrow
{Halting} ($\overline{T}=1.18$)
& 2.02
& 3.73 & 2.55 & \textbf{1.02} & 0.60 & 2.48
& 2.15 & 9.86 & 1.28 & 0.58
& 10.90 & 0.92 & \textbf{1.25} & 4.20
& \textbf{0.00} & 0.44 & \textbf{0.00}\\

\midrule

\tablebaselinerow
Oracle ($\overline{T}=1.08$)
& --
& 3.39 & 2.01 & 1.02 & 0.34 & 1.74 & 1.78 & 8.77
& 0.42 & 0.16 & 8.28 & 0.63 & 1.18 & 3.81
& 0.00 & 0.00 & 0.00\\

\bottomrule
}

\resizebox{\textwidth}{!}{%
\begin{tabular}{
    @{\hspace{4pt}}l*{17}{c}@{\hspace{4pt}}
}
\coreLoopMainTableBody
\end{tabular}%
}
\endgroup
\end{table*}

\subsection{Experimental setup}
\label{ssec:setup}

\noindent\textbf{Data and metrics.}
Broadly following Teffic Audio~\citep{lin2026tefficaudiotellfactfiction},
we train on the official training splits of 14 public datasets:
GigaSpeech, SpeechFake, SpoofCeleb, DFADD, CN-Celeb, MLAAD,
Fake or Real, AISHELL-3, ASVspoof 2019 LA, ASVspoof 5,
LibriSpeech, CodecFake, ADD 2023, and ADD 2022~\citep{
du2024dfadddiffusionflowmatchingbased, GigaSpeech2021,
huang2025speechfake, jung2025spoofcelebspeechdeepfakedetection,
li2021cncelebmultigenrespeakerrecognition,
müller2026mlaadmultilanguageaudioantispoofing,
panayotov2015librispeech, reimao2019dataset,
shi2021aishell3multispeakermandarintts,
wang2024asvspoof5crowdsourcedspeech,
wang2020asvspoof2019largescalepublic,
xie2024codecfakedatasetcountermeasuresuniversally,
yi2023add2023secondaudio, yi2024add2022audiodeep}.
No evaluation or test utterances are used for training.

We evaluate on the 14 public test conditions in Speech DF
Arena~\citep{11345101}, covering ASVspoof, CodecFake, In the Wild,
LibriSeVoc, SONAR, DFADD, Fake or Real, and ADD benchmarks.
We report pooled EER as the primary metric by combining scores from all
conditions, together with their unweighted macro-average EER.
RTF is inference time divided by input duration, computed from the
waveform length at 16\,kHz. Timing uses 4\,s inputs, batch size 1,
FP16 inference, and a single NVIDIA V100 GPU.

\noindent\textbf{Models and optimization.}
The primary model uses a frozen 24-layer W2V-BERT encoder~\cite{w2v-bert}, while scaling
experiments additionally consider its 3-, 6-, and 12-layer variants.
The base detector is trained for 30 epochs using BCE and then frozen.
All adapted models start from the same checkpoint. LoRA uses rank
$r=8$, scaling $\alpha=16$, and dropout $0.05$. The LoRA adapters
and loop-specific modules are trained for 10 epochs using AdamW with
learning rate $10^{-4}$, batch size 32, and BF16 precision; all original
encoder and classifier parameters remain frozen.

\noindent\textbf{Comparisons.}
We compare \method{} with the frozen detector, single-pass
LoRA~\citep{11463170}, naive recurrence, and fixed-depth inference at
$T\in\{1,2,3\}$. Single-pass LoRA uses the same adapter placement,
rank, and training schedule as \method{}, but excludes all recurrent-state
modules. Naive recurrence instead repeatedly applies the shared core
without the complete recurrent interface used by \method{}.

For adaptive inference, all policies use the same trained refinement
model and must exit at $T=3$. Random halting exits independently with
probability $0.5$ at each of the first two passes, giving an expected
depth of $1.75$. Confidence halting exits when the current binary-class
confidence reaches $0.9$. The non-deployable oracle has access to
ground-truth labels and future predictions. Using the pooled-EER
threshold at each pass, it continues only when a later pass can correct
the current error; otherwise, it exits immediately.

\subsection{Main results across test conditions}
\label{ssec:main_results}

Table~\ref{tab:main} shows that \method{} at $T=2$ reduces pooled EER
from 4.85\% to 3.74\%, corresponding to a 22.9\% relative reduction over
the frozen baseline, with improvements on 10 of the 14 test sets. It also
outperforms single-pass LoRA at 4.24\% pooled EER, indicating that LoRA
adaptation alone cannot explain the full gain brought by recurrent
refinement. A third pass yields only a marginal additional reduction in
pooled EER while slightly worsening macro EER, suggesting that most of the
available benefit has already been captured after the first refinement pass
and that further recurrence provides increasingly limited returns.

Across three random seeds (42, 0, 1) using the same frozen checkpoint, fixed T=2, fixed T=3, and adaptive halting obtain pooled EERs of 3.81\ensuremath{\pm}0.07\%, 3.78\ensuremath{\pm}0.08\%, and 3.79\ensuremath{\pm}0.06\% (mean\ensuremath{\pm}std), respectively, confirming stable performance across random seeds.

Adaptive halting preserves the fixed-depth gain while concentrating
computation on a smaller subset of utterances. It obtains 3.73\% pooled EER
at $\overline{T}=1.18$ and an RTF of $2.02\times10^{-2}$. Relative to the
frozen detector, this is a 23.1\% EER reduction for an 18.1\% RTF increase;
relative to fixed $T=2$, it reduces RTF by 44.5\% with essentially unchanged
pooled EER. Confidence and random routing obtain 4.08\% and 4.03\% pooled
EER despite using average depths of 1.06 and 1.75, respectively. Their
different computation budgets but similar degradation indicate that useful
depth selection depends on which samples continue, not simply how many do.
The oracle result of 3.39\% further suggests that routing quality, rather than
the availability of additional refinement alone, remains an important source
of performance headroom.

All evaluation utterances are held out, but the 14 conditions do not share
one uniform notion of novelty. Some isolate generator-level novelty, whereas
others primarily introduce new datasets, channels, or recording conditions.
We therefore interpret the aggregate results as performance across
diverse held-out distributions rather than universal improvement on
unseen generators. Appendix~\ref{app:scope_audit} documents the supported
claim for each condition.

\subsection{Backbone scaling}
\label{ssec:scaling}

\begin{table}[t]
\centering
\caption{Scaling across different frozen encoder depths. Entries report
pooled/macro EER (\%). $\bar{T}$ denotes the average number of refinement
passes used during adaptive inference.}
\label{tab:scaling}
\begingroup
\small
\setlength{\tabcolsep}{0.7pt}
\renewcommand{\arraystretch}{1.12}
\begin{tabular*}{\columnwidth}{
@{\hspace{4pt}\extracolsep{\fill}}cccccc@{\hspace{4pt}}
}
\toprule
Layers & Baseline & $T=2$ &
$T=3$ & Adaptive & \textbf{$\bar{T}$} \\
\midrule
3  & 7.74/5.48 & 6.23/4.42 &
6.26/4.36 & 6.25/4.40 & 1.36 \\
6  & 5.32/3.90 & 4.25/2.93 &
4.19/2.92 & 4.38/3.07 & 1.15 \\
12 & 5.28/3.93 & 4.19/2.72 &
4.17/2.74 & 4.23/3.11 & 1.21 \\
24 & 4.85/2.86 & 3.74/2.47 &
3.69/2.49 & 3.73/2.55 & 1.18 \\
\bottomrule
\end{tabular*}
\endgroup
\end{table}

Table~\ref{tab:scaling} shows that two-pass refinement improves both pooled and macro EER at every encoder depth, reducing pooled EER by 1.07--1.51 percentage points. A third pass yields only small, mixed changes across model sizes, suggesting that early saturation is not specific to the 24-layer encoder. Adaptive inference stays within 0.13 percentage points of fixed $T=2$ in pooled EER while using 32--42.5\% fewer Transformer-core evaluations. Its macro EER is less consistent, particularly at 12 layers (3.11\% versus 2.72\%). Complete condition-level EERs and adaptive exit depths for the 3-, 6-, and 12-layer models are reported in Appendix~\ref{app:scaling_full}.

\section{Refinement and Routing Analysis}
\label{sec:analysis}

We next examine how recurrent refinement changes predictions, which
components make reuse of the frozen encoder effective, why later passes
saturate, and whether the halting policy assigns computation to samples that
can benefit from it.

\begin{table}[t]
\centering
\caption{Mechanism controls and component ablations. Each entry reports
pooled/macro EER (\%). Lower is better. Minimal retains only LoopBridge,
ExitBridge, and loop LoRA.}
\label{tab:ablation}
\begingroup
\small
\setlength{\tabcolsep}{4pt}
\renewcommand{\arraystretch}{1.12}

\begin{tabularx}{\columnwidth}{
    @{\hspace{4pt}}Xcc@{\hspace{4pt}}
}
\toprule
Variant & $T=2$ & $T=3$ \\
\midrule
\tablebaselinerow
Zero-shot naive recurrence    & 10.52/8.02 & 64.68/58.19 \\
\tablebaselinerow
Naive recurrence w/ loop LoRA & 4.57/3.47  & 4.57/3.54 \\
\midrule
\method{} w/o LoopBridge      & 4.56/3.35 & 4.45/3.16 \\
\method{} w/o ExitBridge      & 3.95/2.49 & 3.79/2.44 \\
\method{} w/o loop LoRA       & 4.12/2.80 & 4.08/2.73 \\
\method{} w/o anchor bank     & 3.90/2.46 & 3.85/2.41 \\
\method{} w/o UpdateGate      & 3.78/2.58 & 3.79/2.66 \\
\method{} w/o middle inject.  & 3.87/2.56 & 3.85/2.49 \\
\method{}, 8 anchors          & 3.96/2.42 & 3.94/2.45 \\
\midrule
\tablemethodrow
\method{} (minimal)           & 3.89/2.52 & 3.91/2.58 \\
\tablebestrow
\method{} (full)              & \textbf{3.74/2.47}
                              & \textbf{3.69/2.49} \\
\bottomrule
\end{tabularx}
\endgroup
\end{table}

\subsection{Sources of recurrent refinement gains}
\label{ssec:ablation}

The controls in Table~\ref{tab:ablation} separate useful refinement from
the effects of simply evaluating the frozen encoder again. Direct recurrence
is severely unstable, and loop LoRA alone recovers only part of the lost
compatibility. LoopBridge is the most consequential component: without it,
the $T=2$ pooled EER rises from 3.74\% to 4.56\%, almost matching naive
recurrence with loop LoRA. This result further supports the central premise that the previous endpoint cannot directly serve as a core input without explicit
interface alignment.

ExitBridge and loop LoRA address complementary mismatches. The former
returns the refined state to the feature space expected by the frozen
classifier, whereas the latter adapts the shared encoder computation to the
new recurrent input. Removing either weakens the gain even when the other
modules remain. A minimal system containing these two bridges and loop LoRA
already reaches 3.89\% pooled EER, showing that input alignment, encoder
adaptation, and output alignment account for most of the improvement. The full model is consistently stronger, particularly at $T=3$, suggesting
that the anchor context and state-control modules become increasingly useful
as recurrent states move farther away from the original feed-forward path.

\subsection{Refinement depth and decision transitions}
\label{ssec:depth_dynamics}

\begin{table}[t]
\centering
\caption{Depth extrapolation on the 24-layer model. Results for
$T=1$--$3$ are reproduced from Table~\ref{tab:main}, while $T=4$--$6$
and adaptive halting with $T_{\max}=6$ are obtained from the separate
six-loop evaluation. The halting policy uses an average depth of
$\overline{T}=1.37$. Lower EER is better; the best result in each column
is bolded.}
\label{tab:depth6}
\begingroup
\small
\renewcommand{\arraystretch}{1.10}
\begin{tabular*}{\columnwidth}{
    @{\hspace{4pt}\extracolsep{\fill}}lcc@{\hspace{4pt}}
}
\toprule
Policy & Pooled & Macro \\
\midrule
Fixed $T=1$ & 4.85 & 2.86 \\
Fixed $T=2$ & 3.74 & \textbf{2.47} \\
Fixed $T=3$ & 3.69 & 2.49 \\
\midrule
Fixed $T=4$ & \textbf{3.65} & 2.51 \\
Fixed $T=5$ & 3.66 & 2.58 \\
Fixed $T=6$ & 3.67 & 2.60 \\
Halting ($T_{\max}=6$) & 3.71 & 2.50 \\
\bottomrule
\end{tabular*}
\endgroup
\end{table}

Table~\ref{tab:depth6} extends the fixed-depth trajectory beyond the training
range. The large gain is concentrated in the first refinement pass; further
unrolling changes pooled EER only marginally, while macro EER gradually
worsens. The lowest pooled value occurs at $T=4$, but the subsequent reversal
shows that additional depth does not produce a stable improvement. The
$T_{\max}=6$ halting result remains close to this plateau while using only
1.37 passes on average, reinforcing that the useful computation is sparse.
Because $T=1$--$3$ reproduce the main experiment and the deeper rows come
from a separate six-loop evaluation, the cross-boundary differences are
descriptive rather than paired estimates; Appendix~\ref{app:depth_protocol}
details this protocol and reports the complete condition-level results.

\begin{figure}[t]
  \centering
  \includegraphics[width=0.95\columnwidth]{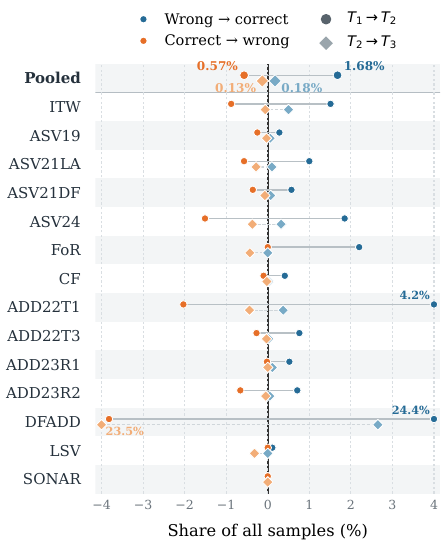}
    \caption{Sample-level error corrections and regressions across successive
refinement passes of the 24-layer model, evaluated on pooled data and
14 test sets.}
  \label{fig:loop_correction}
\end{figure}

Figure~\ref{fig:loop_correction} reveals what the aggregate depth curve hides.
The first transition corrects substantially more samples than it harms
(1.68\% versus 0.57\% of the pooled set), explaining the clear gain at
$T=2$. By the next transition, both events are rare and nearly balanced
(0.18\% versus 0.13\%). Refinement therefore does not cease to change
predictions; rather, the remaining changes increasingly exchange one set of
errors for another. The balance also varies across test conditions, so a
single fixed depth cannot be optimal for every utterance. This transition
view motivates both the internal-state analysis below and sample-wise
halting.

\begin{figure}[t]
  \centering
  \includegraphics[width=0.95\columnwidth]{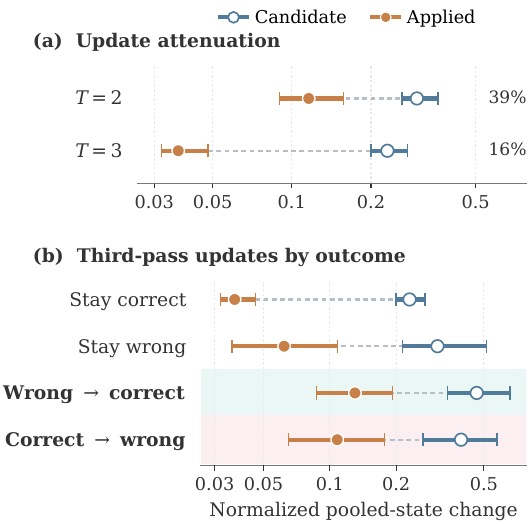}
  \caption{Hidden-state update attenuation under fixed $T=3$ inference.
  Points and horizontal bars indicate medians and interquartile ranges over the pooled evaluation set. Candidate $\delta$ is the proposed
  state change before the UpdateGate, whereas applied $\delta$ is the
  realized change after gating. (a) Update magnitude by pass; percentages
  give the ratio of applied to candidate medians. (b) Third-pass updates
  grouped by the $T=2\rightarrow T=3$ prediction transition.}
  \label{fig:hidden_gate}
\end{figure}

\subsection{Why does refinement saturate?}
\label{ssec:saturation}

The third pass changes pooled EER only marginally. To distinguish a nearly
static hidden state from an attenuated update, we trace all pooled
evaluation utterances with fixed $T=3$ and halting disabled. For $T\geq2$, we measure the candidate and applied changes
relative to the previous pooled representation at each refinement pass:
\begin{equation}
\begin{aligned}
\delta_t^{\mathrm{cand}}
&=\frac{\|\operatorname{pool}(\cstate_t)-
\operatorname{pool}(\s_{t-1})\|_2}
{\|\operatorname{pool}(\s_{t-1})\|_2+\epsilon},\\
\delta_t^{\mathrm{app}}
&=\frac{\|\operatorname{pool}(\s_t)-
\operatorname{pool}(\s_{t-1})\|_2}
{\|\operatorname{pool}(\s_{t-1})\|_2+\epsilon}.
\end{aligned}
\label{eq:hidden_delta}
\end{equation}

Figure~\ref{fig:hidden_gate}(a) distinguishes a dormant representation from
an actively proposed but suppressed update. The candidate change decreases
only moderately from the second to the third pass, whereas the applied
change contracts from 0.116 to 0.037. The applied-to-candidate ratio therefore
falls from 39\% to 16\%, accompanied by a smaller gate and logit change. The
third pass is not a fixed point: the shared core still produces a distinct
state, but the recurrent controller allows much less of that state to alter
the accumulated representation. This attenuation is also supported by our
ablation: removing the UpdateGate slightly degrades performance at $T=3$,
suggesting that controlling update magnitude is beneficial and can suppress
some harmful late-pass updates.

The outcome-conditioned view in Figure~\ref{fig:hidden_gate}(b) adds a second
part to this explanation. Samples whose decisions change in either direction
receive larger candidate and applied updates than stable samples. However, corrective and regressive transitions have similar gate
medians despite their opposite prediction outcomes. These transitions
are nearly balanced at the third pass. Together with the smaller
applied updates, this characterizes limited additional benefit
despite continued changes in recurrent states. Full distributions
are reported in Appendix~\ref{app:hidden_gate}.

Core input--output heatmaps show similar channel--time patterns across
successive passes, with weaker third-pass changes and a smaller fraction
of the input change reaching the core output. On the selected diagnostic
samples, these patterns complement the aggregate state-change statistics
and provide a visual account of later-pass attenuation
(Appendix~\ref{app:core_heatmap}).

\subsection{Does adaptive halting allocate computation effectively?}
\label{ssec:routing_analysis}

Table~\ref{tab:routing_compact} tests whether the learned policy saves
computation merely by exiting early. Recoverable first-pass errors are rare,
but learned halting continues 77.10\% of them, substantially more than the
random and confidence policies, while producing the lowest premature-exit
rate. At the same time, 98.57\% of its first-pass exits are already correct.
The policy therefore combines two behaviors: it removes computation from
easy utterances and preserves more of the small subset for which another
pass can repair the decision.

The halting target is broader than binary error correction. It asks whether
a future pass can reduce BCE, even when the predicted class does not change.
Such opportunities occur for 8.44\% of first-pass samples, compared with
only 1.25\% that are decision-level recoverable errors, and learned halting
recalls 72.53\% of the BCE opportunities. This alignment with the training
target explains why the policy continues more samples than a pure error
detector would require. The complete metric definitions, additional harm
and exit-composition statistics, and per-condition variation are provided
in Appendix~\ref{app:routing_quality}.

\begin{table}[t]
\centering
\caption{Compact pooled routing diagnostics at a fixed $0.5$ decision
threshold. Premature denotes an exit on a first-pass error that a later pass
can correct; recoverable recall measures how often such errors continue.}
\label{tab:routing_compact}
\begingroup
\small
\setlength{\tabcolsep}{1.2pt}
\renewcommand{\arraystretch}{1.12}
\begin{tabular*}{\columnwidth}{
@{\hspace{4pt}\extracolsep{\fill}}lccc@{\hspace{4pt}}
}
\toprule
Policy & $\bar{T}$ & Premature/all & Recoverable recall \\
\midrule
Learned halting & 1.18 & \textbf{0.30\%} & \textbf{77.10\%} \\
Random halting & 1.75 & 0.68\% & 48.58\% \\
Confidence@0.9 & 1.06 & 0.66\% & 48.58\% \\
\bottomrule
\end{tabular*}
\endgroup
\end{table}

\section{Conclusion}
\label{sec:conclusion}

We presented \method{}, which enables recurrent refinement in a frozen
audio deepfake detector through loop-specific LoRA and lightweight state
control. Across 14 diverse test sets, recurrent processing improves
over single-pass inference, with one additional pass capturing most of the
gain, while adaptive halting achieves comparable performance with
substantially fewer shared encoder-core evaluations. Our analyses show smaller applied updates and nearly balanced
corrections and regressions at later passes, despite continued
changes in recurrent states. Ablations highlight the importance
of recurrent-input alignment, with UpdateGate providing a modest
additional benefit. Learned halting also preferentially exits
samples that are already reliably classified while preserving recoverable
errors for further refinement. Overall, \method{} shows that selective
recurrent computation can improve an already-trained detector without
updating its original parameters.

\section*{Limitations}
Our evaluation focuses on audio deepfake detection with a fixed collection
of training and test datasets. The results do not establish that the same
refinement design generalizes to other speech tasks.
The main evaluation uses at most three passes. A post-hoc unrolling
diagnostic extends inference to six passes and shows saturation, but does
not replace a controlled training study of deeper recurrence.
The small additional gains from later passes are associated with gate
attenuation and offsetting prediction flips, while the causal effect of
changing the gate remains unresolved.
Refinement can also overturn correct predictions, and its benefits vary
across test conditions.
The halting head is trained using labels from the original training data,
so its routing behavior under further distribution shifts remains to be
studied.
Reported timing uses a single V100 GPU, batch size one, and four-second
inputs; deployment costs may differ with hardware, batching, and utterance
duration.
The multi-seed results reuse one frozen detector checkpoint and therefore
do not measure variation from independently training the base detector.

\bibliography{refs}

\appendix
\raggedbottom
\setlength{\textfloatsep}{9pt plus 2pt minus 2pt}
\setlength{\dbltextfloatsep}{10pt plus 2pt minus 2pt}

\noindent{\large\bfseries Appendix overview.}\par
\smallskip
Appendix~\ref{app:module_impl} describes the recurrent refinement
modules, implementation settings, and computational budget. Appendix~\ref{app:depth_protocol}
then documents the six-pass extrapolation protocol and its condition-level
behavior, and Appendix~\ref{app:scaling_full} expands the backbone-scaling
results. Appendices~\ref{app:hidden_gate} and~\ref{app:core_heatmap} detail
the state-update analyses used to explain later-pass saturation.
Appendix~\ref{app:routing_quality} defines the routing diagnostics and reports
their pooled and condition-level behavior, while Appendix~\ref{app:scope_audit}
audits which forms of evaluation novelty are supported by the released
metadata. Appendix~\ref{app:ethical} discusses ethical considerations, including data use,
potential risks, and AI assistance.
Throughout the appendix, shaded table headers identify comparable
blocks, gray denotes a baseline or reference, blue denotes recurrent-model
results, and green highlights the selected or best-performing configuration.

\section{Implementation details}
\label{app:module_impl}

This section provides additional implementation details for the recurrent
modules used in the 24-layer configuration. We follow the notation and
formulation in the main text. The encoder hidden size is $H=1024$, the
anchor-context dimension is $E=256$, and the bridge bottleneck dimension is
256. Unless otherwise specified, lightweight MLPs use a hidden width of 256
and dropout 0.1. All recurrent modules are bypassed on the first pass.

\subsection{Anchor bank and AnchorReader}
\label{app:anchor_reader}

\paragraph{Anchor bank.}
The 24-layer model uses the first-pass Prelude endpoint and encoder
states at layers $\{6,12,18,24\}$ as anchor sources. Each source is
independently normalized and projected from 1024 to 256 dimensions.
The five anchor sequences are computed once and remain fixed during
refinement. Anchor-branch gradients are scaled by 0.2 without changing
forward values.

\paragraph{AnchorReader.}
For $t\geq2$, a mask-aware temporal mean of the previous endpoint is
combined with the step embedding to predict utterance-level source
weights, shared across frames:
\begin{equation}
\boldsymbol{\alpha}_t
=
\operatorname{softmax}\left(R([\mathbf{q}_t;\tau_t])\right),
\label{eq:app_anchor_weights}
\end{equation}
\begin{equation}
\mathbf{m}_t(j)
=
\sum_{n=1}^{N}\alpha_{t,n}
\widetilde{\mathbf{a}}^{(n)}(j),
\qquad N=5.
\label{eq:app_anchor_mix}
\end{equation}
Here, $R$ is an $(H+S)\rightarrow256\rightarrow N$ MLP
with SiLU activation and dropout 0.1.

A frame-wise gate compares the previous state, projected into anchor
space, with $\mathbf{m}_t(j)$. Its input concatenates the state,
anchor, absolute difference, element-wise product, and step embedding.
A $(4E+S)\rightarrow256\rightarrow1$ MLP produces:
\begin{equation}
\begin{aligned}
\rho_t(j)&=\sigma\left(G(\mathbf{u}_t(j))\right),\\
\e_t(j)&=\rho_t(j)\mathbf{m}_t(j).
\end{aligned}
\label{eq:app_anchor_gate}
\end{equation}
Padding positions are masked after gating. Source selection is thus
utterance-level, while anchor retention varies across frames.

\subsection{LoopBridge}
\label{app:loopbridge_impl}

LoopBridge constructs the recurrent input in
Eq.~\eqref{eq:loop_bridge}. It compares the previous endpoint with
the fixed Prelude anchor through a
$1024\rightarrow256\rightarrow1024$ state adapter.
The anchor-context adapter uses $256\rightarrow256\rightarrow1024$,
and a linear projection maps the step embedding to the encoder space.

Both adapters use SiLU, with dropout 0.1 on the state path.
Their output projections and the step projection are zero initialized,
so the recurrent input initially matches the Prelude representation.
Removing anchor context omits its term in
Eq.~\eqref{eq:loop_bridge}; ablating LoopBridge directly reuses the
previous endpoint as the core input.

\subsection{Loop-specific LoRA and anchor injection}
\label{app:lora_injection_impl}

LoRA is applied to the attention $q/k/v/o$ projections and FFN
linear layers of the shared core, with rank $r=8$, $\alpha=16$,
and dropout 0.05. Its factors are shared across recurrent passes,
while a learned step-dependent scalar modulates their contribution.
The output projection is zero initialized, and LoRA is disabled
on the first pass.

The routed anchor context is also injected after layer 12 in the
24-layer model. A zero-initialized linear projection maps it from
256 to 1024 dimensions, with a learned step-dependent scale.
This injection is active only during refinement.

\subsection{UpdateGate}

UpdateGate implements the interpolation in Eq.~(2).
Mask-aware temporal means of the previous endpoint and candidate
state, their difference, the anchor summary, and the step embedding
are concatenated and processed by a
$(3H+E+S)\rightarrow256\rightarrow H$ MLP.
For $H=1024$, $E=256$, and $S=64$, the input dimension is 3392.
The MLP uses SiLU, dropout 0.1, and an output sigmoid;
unavailable anchor summaries are replaced with zeros.

The gate provides one value per hidden channel and utterance,
broadcast across frames. Padding is excluded from temporal summaries.
The final-layer bias is initialized to $-1.73$
($\sigma(-1.73)\approx0.15$), while weights retain their default
initialization. This trainable bias favors conservative initial
updates while preserving input-dependent gates.
Without UpdateGate, the candidate directly replaces the previous state.

\subsection{ExitBridge}
\label{app:exitbridge_impl}

For terminal depths greater than one, ExitBridge aligns the recurrent
endpoint with the first-pass output space before the frozen classifier.
The normalized difference from the first-pass endpoint is concatenated
with the step embedding and passed through an
$(H+S)\rightarrow256\rightarrow H$ residual adapter.
It uses SiLU, dropout 0.1, and a zero-initialized final projection,
so ExitBridge initially implements an identity mapping.

\subsection{Adaptive halting head}
\label{app:halting_impl}

The halting head is trained with the recurrent refiner frozen.
At each non-terminal depth, it uses the current and previous states
and their classification logits. Mask-aware temporal means and
standard deviations summarize the current state and its change.
These statistics are projected from $2H$ to 256 and 128 dimensions,
respectively, each followed by GELU activation.

Five scalar features capture prediction confidence and state change:
\begin{equation}
\begin{aligned}
c_t^{\mathrm{sim}}
&=\cos(\boldsymbol{\mu}_t,\boldsymbol{\mu}_{t-1}),\\
r_t
&=\frac{\|\s_t-\s_{t-1}\|_2}
        {\|\s_{t-1}\|_2+10^{-8}},
\end{aligned}
\end{equation}
\begin{equation}
\mathbf{v}_t=
[z_t,\ |z_t|,\ z_t-z_{t-1},\
c_t^{\mathrm{sim}},\ r_t].
\label{eq:app_halting_scalars}
\end{equation}
Here, $\boldsymbol{\mu}_t$ is the mask-aware temporal mean.
The projected statistics, scalar features, and a 16-dimensional
step embedding form a 405-dimensional input.
After LayerNorm, a $405\rightarrow256\rightarrow64\rightarrow1$
MLP with GELU and dropout 0.1 predicts the continuation logit $a_t$.
Targets, training objective, and inference follow
Section~\ref{ssec:training_halting}.

\begin{table}[t]
\centering
\caption{Trainable parameter overhead of the main recurrent components in
the 24-layer configuration. LoRA is reported separately because it is
attached directly to frozen core linear layers.}
\label{tab:module_params}
\begingroup
\small
\setlength{\tabcolsep}{0pt}
\renewcommand{\arraystretch}{1.08}
\begin{tabular*}{\columnwidth}
{@{\hspace{4pt}}l@{\extracolsep{\fill}}r@{\hspace{4pt}}}
\toprule
\textbf{Component} & \textbf{Parameters (M)} \\
\midrule
Anchor projections & 1.32 \\
AnchorReader & 0.83 \\
LoopBridge & 0.92 \\
UpdateGate & 1.13 \\
ExitBridge & 0.54 \\
\midrule
Listed recurrent modules & 4.75 \\
Loop-specific LoRA (24 layers) & 5.51 \\
Halting head & 0.91 \\
\bottomrule
\end{tabular*}
\endgroup
\end{table}

\subsection{Implementation details and computational budget}
\label{app:implementation_overhead}

\paragraph{Software and model initialization.}
Experiments use Python 3.10.16, PyTorch 2.6.0 with CUDA 12.4,
Transformers 4.45.0, and torchaudio 2.6.0.
We load the pretrained w2v-BERT 2.0 encoder and its
SeamlessM4T feature extractor through Transformers.
The encoder contains 24 Conformer layers with a hidden size of 1024.
The recurrent model is initialized from the trained detector checkpoint.
Training uses automatic mixed precision with bfloat16.

\paragraph{Audio preprocessing and feature extraction.}
Audio is loaded and resampled using torchaudio, converted to mono
by averaging channels, and resampled to 16\,kHz.
Training inputs contain 64,600 samples (approximately 4.04 seconds),
while validation and test inputs contain 64,000 samples (4 seconds).
Long utterances are randomly cropped during training and validation;
testing retains the first 64,000 samples.
Short utterances are zero-padded with probability 0.5, with padding
randomly distributed between the two sides; otherwise, they are
repeated and truncated to the target length.
This short-utterance handling is used in all splits.

Waveforms with an absolute peak above one are divided by that peak.
The feature extractor scales waveforms by \(215\) and extracts
80-dimensional log-mel features using a 400-sample window,
a 160-sample hop, pre-emphasis of 0.97, and DC removal.
Each mel channel is then normalized to zero mean and unit variance.

\paragraph{Masking.}
All recurrent utterance-level summaries use the encoder attention
mask so that padding frames do not contribute.
AnchorReader also sets its frame-wise gate to zero at padded positions.
The same encoder mask is retained when recurrent states are processed
by the shared core.

\paragraph{Evaluation implementation.}
EER is computed using the ROC curve implementation in scikit-learn 1.6.1.
We select the ROC point minimizing the absolute difference between
the false-negative and false-positive rates and report their average.
Pooled EER is computed from concatenated scores across all 14
evaluation conditions; macro EER is the unweighted mean of their
condition-specific EERs.

\paragraph{Parameter overhead and training budget.}
Table~\ref{tab:module_params} summarizes the main trainable parameter
overhead of the 24-layer configuration, excluding the frozen detector
and original classifier.
The listed recurrent modules contain approximately 4.75M trainable
parameters, while the LoRA adapters contribute another 5.51M.
The separately trained halting head contains 0.91M parameters,
bringing the total additional trainable parameter count to
approximately 11.17M.
All original detector parameters remain frozen.
Training uses four NVIDIA A100 GPUs, with a total training budget
of approximately 20 GPU hours.

\begin{table*}[!t]
\centering
\caption{Condition-level EER (\%) under depth extrapolation.
$T=1$--$3$ reproduce the main evaluation, while $T=4$--$6$,
adaptive halting, and the oracle use six-pass unrolling.
$\bar{T}$ denotes the average exit depth. Oracle is reported only as a
non-deployable reference and is excluded from the boldface comparison.
Lower is better; the best non-oracle EER in each row is bolded.}
\label{tab:depth6_full}

\begingroup
\footnotesize

\setlength{\tabcolsep}{2.5pt}
\renewcommand{\arraystretch}{1.08}

\begin{tabularx}{\textwidth}{
@{\hspace{4pt}}
l
>{\centering\arraybackslash\columncolor{tblBaseline}}X
>{\centering\arraybackslash\columncolor{tblBaseline}}X
>{\centering\arraybackslash\columncolor{tblBaseline}}X
>{\centering\arraybackslash\columncolor{tblCore}}X
>{\centering\arraybackslash\columncolor{tblCore}}X
>{\centering\arraybackslash\columncolor{tblCore}}X
>{\centering\arraybackslash\columncolor{tblHalting}}X
>{\centering\arraybackslash\columncolor{tblHalting}}X
|
>{\centering\arraybackslash}X
>{\centering\arraybackslash}X
@{\hspace{4pt}}
}
\toprule

\multirow{2}{*}{\textbf{Condition}}
& \multicolumn{3}{c}{\textbf{Training range}}
& \multicolumn{3}{c}{\textbf{Extrapolated}}
& \multicolumn{2}{c}{\textbf{Six-pass halting}}
& \multicolumn{2}{c}{\textbf{Six-pass oracle}} \\

\cmidrule(lr){2-4}
\cmidrule(lr){5-7}
\cmidrule(lr){8-9}
\cmidrule(lr){10-11}

& \textbf{$T=1$}
& \textbf{$T=2$}
& \textbf{$T=3$}
& \textbf{$T=4$}
& \textbf{$T=5$}
& \textbf{$T=6$}
& \textbf{EER}
& \textbf{$\bar{T}$}
& \textbf{EER}
& \textbf{$\bar{T}$} \\

\midrule

In the Wild
& 1.88 & 1.66 & \textbf{1.24}
& 1.56 & 1.56 & 1.56
& 1.28 & 1.27
& 0.49 & 1.04 \\

ASVspoof 2019
& \textbf{0.34} & 0.63 & 0.62
& 0.48 & 0.57 & 0.48
& 0.49 & 1.12
& 0.32 & 1.02 \\

ASVspoof 2021 LA
& 2.73 & \textbf{2.26} & 2.39
& 2.62 & 2.53 & 2.54
& 2.52 & 1.34
& 1.79 & 1.10 \\

ASVspoof 2021 DF
& 2.69 & 2.18 & 2.18
& 2.14 & 2.18 & 2.17
& \textbf{2.02} & 1.16
& 1.59 & 1.09 \\

ASVspoof 5
& 10.43 & \textbf{9.53} & 9.56
& 9.61 & 9.61 & 9.56
& 9.99 & 1.40
& 8.58 & 1.41 \\

Fake or Real
& 2.15 & 0.86 & 1.28
& 0.86 & \textbf{0.42} & \textbf{0.42}
& \textbf{0.42} & 1.60
& 0.00 & 1.06 \\

CodecFake
& 0.22 & 0.21 & 0.22
& \textbf{0.16} & 0.17 & 0.19
& 0.54 & 1.19
& 0.09 & 1.01 \\

ADD 2022 T1
& 12.45 & \textbf{10.55} & 10.60
& 10.66 & 10.78 & 10.69
& 11.01 & 2.33
& 7.96 & 1.45 \\

ADD 2022 T3
& 0.90 & 0.81 & 0.92
& 0.80 & 0.73 & 0.89
& \textbf{0.69} & 1.20
& 0.33 & 1.03 \\

ADD 2023 R1
& 1.63 & 1.36 & 1.25
& 1.14 & \textbf{0.99} & 1.02
& 1.06 & 1.17
& 0.99 & 1.06 \\

ADD 2023 R2
& 4.51 & 4.20 & 4.20
& \textbf{4.11} & 4.21 & 4.18
& 4.28 & 1.36
& 3.84 & 1.18 \\

DFADD
& \textbf{0.00} & \textbf{0.00} & \textbf{0.00}
& \textbf{0.00} & \textbf{0.00} & \textbf{0.00}
& \textbf{0.00} & 1.00
& 0.00 & 1.01 \\

LibriSeVoc
& \textbf{0.13} & 0.38 & 0.44
& 1.01 & 2.34 & 2.71
& 0.69 & 2.47
& 0.13 & 1.03 \\

SONAR
& \textbf{0.00} & \textbf{0.00} & \textbf{0.00}
& \textbf{0.00} & \textbf{0.00} & \textbf{0.00}
& \textbf{0.00} & 1.33
& 0.00 & 1.01 \\

\midrule

\textbf{Pooled}
& 4.85 & 3.74 & 3.69
& \textbf{3.65} & 3.66 & 3.67
& 3.71 & 1.37
& 3.20 & 1.18 \\

\textbf{Macro}
& 2.86 & \textbf{2.47} & 2.49
& 2.51 & 2.58 & 2.60
& 2.50 & --
& 1.87 & -- \\

\bottomrule
\end{tabularx}

\endgroup
\end{table*}

\section{Protocol for six-pass depth extrapolation}
\label{app:depth_protocol}

\subsection{Evaluation design and reporting conventions}

The standard refiner is trained by sampling terminal depths from
$\{2,3\}$, and all primary comparisons in Table~\ref{tab:main} use
$T_{\max}=3$. Table~\ref{tab:depth6} is a post-hoc stress test of the same
refinement mechanism beyond that training range. No deeper supervision is
introduced: the shared core and refinement modules are repeatedly applied up
to six passes, and the $T_{\max}=6$ halting configuration is evaluated on the
resulting trajectories. The experiment therefore asks whether the learned
update remains useful when extrapolated, not whether a model explicitly
trained for six passes could perform better.

The $T=1$--$3$ rows in Table~\ref{tab:depth6} reproduce the primary
results in Table~\ref{tab:main}, while $T=4$--$6$ come from a separate
six-pass evaluation. For each new fixed depth, pooled EER is computed
from concatenated scores across all 14 conditions, and macro EER averages
their condition-specific EERs. The halting row uses the same six-pass
evaluation and reports the policy's actual selected depths. Because the
two parts originate from separate runs, differences across the
$T=3$/$T=4$ boundary only characterize the plateau. Comparisons within
$T=4$--$6$ provide clearer evidence that further unrolling does not yield
a sustained gain.

Here, $T$ counts complete evaluations of the shared encoder path:
$T=1$ is the unchanged first-pass detector, and $T=2$ adds one recurrent
refinement step. All depths reuse the same trained parameters;
extrapolation changes only their number of applications.
Condition-level EER uses a threshold estimated within each condition,
whereas pooled EER concatenates all scores before estimating one
threshold. Macro EER is the unweighted mean of the 14 condition-level
EERs. Thus, pooled EER reflects sample-weighted performance, while macro EER
weights all conditions equally.

\subsection{Condition-level depth behavior}

Table~\ref{tab:depth6_full} shows that the pooled plateau masks substantially
different condition-level trajectories. Additional passes beyond $T=3$
retain or slightly improve performance on ASVspoof 2019, Fake or Real,
CodecFake, and ADD 2023 R1, but the preferred depth varies across
conditions. For example, CodecFake reaches its lowest fixed-depth EER at
$T=4$, ADD 2023 R1 continues improving until $T=5$, whereas LibriSeVoc
deteriorates markedly under deeper unrolling. DFADD and SONAR remain
unchanged because their EER is already zero across all fixed depths.
Consequently, the gradual increase in macro EER after $T=3$ is driven by
concentrated regressions on a small subset of conditions rather than a
uniform degradation across the benchmark. This heterogeneity also explains
why no single deeper fixed depth consistently improves over $T=2$ or
$T=3$.

The six-pass oracle provides a complementary reference for the potential of
sample-dependent depth selection. By selecting the earliest depth that
yields a correct prediction, it reaches 3.20\% pooled EER with an average
depth of only 1.18, compared with the best fixed-depth pooled EER of 3.65\%.
Its condition-level results similarly show that useful refinement depth can
vary within a test condition rather than only across conditions. For
example, the oracle reaches 1.79\% on ASVspoof 2021 LA, 1.59\% on
ASVspoof 2021 DF, and 7.96\% on ADD 2022 T1, despite requiring little more
than one pass on average. As the oracle is non-deployable and is computed
from the separate six-pass trajectories, these values are used only as a
reference rather than as part of the fixed-depth comparison. Nevertheless,
they indicate that the saturation of fixed-depth inference does not imply
that later refinement is universally useless; rather, its remaining benefit
is sparse and sample dependent.

\subsection{Prediction transitions and routed depth}

\begin{table}[!t]
\centering
\caption{Pooled prediction changes under six-pass extrapolation. Entries are
percentages of all aligned evaluation samples. The first two transitions
follow the primary analysis; later transitions come from the extrapolation
run.}
\label{tab:depth6_transitions}
\begingroup
\small
\setlength{\tabcolsep}{4pt}
\renewcommand{\arraystretch}{1.10}
\begin{tabular*}{\columnwidth}{
@{\hspace{4pt}\extracolsep{\fill}}lcc@{\hspace{4pt}}
}
\toprule
Transition & Wrong $\rightarrow$ correct &
Correct $\rightarrow$ wrong \\
\midrule
$T=1\rightarrow2$ & 1.68 & 0.57 \\
$T=2\rightarrow3$ & 0.18 & 0.13 \\
$T=3\rightarrow4$ & 0.20 & 0.18 \\
$T=4\rightarrow5$ & 0.07 & 0.09 \\
$T=5\rightarrow6$ & 0.03 & 0.04 \\
\bottomrule
\end{tabular*}
\endgroup
\end{table}

The transition statistics in Table~\ref{tab:depth6_transitions} reinforce
this interpretation at the sample level. The first refinement pass produces
a clear positive correction margin, correcting 1.68\% of all samples while
harming 0.57\%. By $T=2\rightarrow3$, both events become much rarer
(0.18\% versus 0.13\%), consistent with the sharp saturation observed in
the aggregate EER. The $T=3\rightarrow4$ transition still retains a small
positive margin, which coincides with the minimum pooled EER at $T=4$.
After that point, regressions slightly exceed corrections: only
0.07\%/0.03\% of samples are corrected at the next two transitions,
compared with 0.09\%/0.04\% that are harmed. Deeper recurrence still modifies a few decisions, but these changes
no longer yield a consistently positive net effect.

The six-pass halting policy avoids much of this unnecessary computation,
achieving 3.71\% pooled EER with only 1.37 passes on average. Its allocation
is highly condition dependent: the mean depth remains approximately one on
DFADD, but increases to 2.33 on ADD 2022 T1 and 2.47 on LibriSeVoc. The
latter case also illustrates that assigning more computation does not by
itself guarantee improvement when refinement is extrapolated beyond its
training range. The oracle further clarifies the remaining routing
headroom: it reduces pooled EER to 3.20\% while using an even smaller
average depth of 1.18. Thus, the gap between learned halting and the oracle
is not primarily a need for more recurrent computation, but for more
accurate identification of the relatively rare samples and depths for
which additional refinement is beneficial.

\section{Complete backbone-scaling results}
\label{app:scaling_full}

\subsection{Reporting conventions}

Tables~\ref{tab:scaling_3_full}--\ref{tab:scaling_12_full} report the
condition-level results underlying the compact scaling comparison in
Table~\ref{tab:scaling}. Here, fixed $T=1$ exactly reproduces the corresponding frozen baseline in Table~\ref{tab:scaling}, as all refinement modules are bypassed during the first pass. The final row gives
pooled/macro EER; entries that also appear in the main text are reproduced
verbatim. For adaptive inference, $\bar{T}$ is computed separately for each
condition, while the final value is averaged over all evaluation samples.

Within each table, fixed-depth columns use the same number of passes for
every utterance, while the adaptive column uses an utterance-specific depth
and should be read alongside $\bar{T}$. A larger $\bar{T}$ indicates more
frequent refinement, but not necessarily a larger EER reduction, since
the usefulness and direction of the updates also matter. Per-condition
rows use local EER thresholds; pooled EER in the final row is recomputed
from concatenated scores rather than averaged across conditions.

\subsection{Results by encoder depth}

\noindent\textbf{3-layer encoder.}
Table~\ref{tab:scaling_3_full} shows that even the shallowest encoder
benefits from refinement, with two-pass gains on Fake or Real
(8.64\% to 3.45\%) and ASVspoof 2021 LA (11.25\% to 8.99\%).
A third pass has mixed effects: it improves Fake or Real to 2.19\%
but worsens ASVspoof 2021 DF (2.69\% to 3.14\%).
Adaptive inference achieves 6.25\% pooled EER at $\bar{T}=1.36$.

\noindent\textbf{6-layer encoder.}
Table~\ref{tab:scaling_6_full} shows substantial two-pass gains on
Fake or Real (5.60\% to 0.44\%) and CodecFake (1.33\% to 0.18\%).
Adaptive inference reduces average depth to $\bar{T}=1.15$,
with pooled/macro EER of 4.38/3.07\% versus 4.25/2.93\%
at fixed $T=2$, trading a modest performance loss for fewer passes.

\noindent\textbf{12-layer encoder.}
Table~\ref{tab:scaling_12_full} shows refinement benefits at greater
depth, with two-pass gains on Fake or Real (3.02\% to 0.44\%)
and ADD 2022 T1 (14.51\% to 11.24\%).
Adaptive inference improves In the Wild over fixed $T=2$
(1.19\% versus 1.51\%) but worsens CodecFake
(1.41\% versus 0.24\%), achieving 4.23\% pooled EER
at $\bar{T}=1.21$.

\begin{table*}[!t]
\centering
\caption{Complete 3-layer condition-level results. EERs are in percent.}
\label{tab:scaling_3_full}
\begingroup
\small
\setlength{\tabcolsep}{4pt}
\renewcommand{\arraystretch}{1.08}

\begin{tabularx}{\textwidth}{
@{\hspace{4pt}}
l
>{\centering\arraybackslash}X
>{\centering\arraybackslash}X
>{\centering\arraybackslash}X
>{\centering\arraybackslash}X
>{\centering\arraybackslash}X
@{\hspace{4pt}}
}
\toprule
\textbf{Condition} & \textbf{Fixed $T=1$} & \textbf{Fixed $T=2$} &
\textbf{Fixed $T=3$} &
\cellcolor{tblHalting}\textbf{Adaptive} &
\cellcolor{tblHalting}\textbf{$\bar{T}$} \\
\midrule

In the Wild
& 2.89 & \textbf{2.07} & 2.29
& \cellcolor{tblHalting}2.20
& \cellcolor{tblHalting}1.27 \\

ASVspoof 2019
& 4.10 & 3.02 & \textbf{2.73}
& \cellcolor{tblHalting}2.91
& \cellcolor{tblHalting}1.22 \\

ASVspoof 2021 LA
& 11.25 & 8.99 & 8.81
& \cellcolor{tblHalting}\textbf{8.80}
& \cellcolor{tblHalting}1.53 \\

ASVspoof 2021 DF
& \textbf{2.55} & 2.69 & 3.14
& \cellcolor{tblHalting}3.10
& \cellcolor{tblHalting}1.19 \\

ASVspoof 5
& 15.95 & \textbf{14.67} & 14.88
& \cellcolor{tblHalting}14.85
& \cellcolor{tblHalting}1.55 \\

Fake or Real
& 8.64 & 3.45 & 2.19
& \cellcolor{tblHalting}\textbf{1.33}
& \cellcolor{tblHalting}1.71 \\

CodecFake
& 1.36 & 0.30 & \textbf{0.28}
& \cellcolor{tblHalting}0.60
& \cellcolor{tblHalting}1.18 \\

ADD 2022 T1
& 16.15 & \textbf{14.63} & 15.00
& \cellcolor{tblHalting}14.91
& \cellcolor{tblHalting}1.93 \\

ADD 2022 T3
& 1.63 & 1.37 & 1.25
& \cellcolor{tblHalting}\textbf{1.15}
& \cellcolor{tblHalting}1.17 \\

ADD 2023 R1
& 3.07 & 2.62 & \textbf{2.47}
& \cellcolor{tblHalting}2.50
& \cellcolor{tblHalting}1.15 \\

ADD 2023 R2
& 7.25 & 6.85 & 6.74
& \cellcolor{tblHalting}\textbf{6.72}
& \cellcolor{tblHalting}1.29 \\

DFADD
& \textbf{0.00} & \textbf{0.00} & \textbf{0.00}
& \cellcolor{tblHalting}\textbf{0.00}
& \cellcolor{tblHalting}1.09 \\

LibriSeVoc
& \textbf{0.06} & \textbf{0.06} & \textbf{0.06}
& \cellcolor{tblHalting}\textbf{0.06}
& \cellcolor{tblHalting}1.20 \\

SONAR
& 1.82 & \textbf{1.22} & \textbf{1.22}
& \cellcolor{tblHalting}2.44
& \cellcolor{tblHalting}1.16 \\

\midrule

\textbf{Pooled/Macro}
& 7.74/5.48
& \textbf{6.23}/4.42
& 6.26/\textbf{4.36}
& \cellcolor{tblHalting}6.25/4.40
& \cellcolor{tblHalting}1.36 \\

\bottomrule
\end{tabularx}
\endgroup
\end{table*}

\begin{table*}[!t]
\centering
\caption{Complete 6-layer condition-level results. EERs are in percent.}
\label{tab:scaling_6_full}
\begingroup
\small
\setlength{\tabcolsep}{4pt}
\renewcommand{\arraystretch}{1.08}

\begin{tabularx}{\textwidth}{
@{\hspace{4pt}}
l
>{\centering\arraybackslash}X
>{\centering\arraybackslash}X
>{\centering\arraybackslash}X
>{\centering\arraybackslash}X
>{\centering\arraybackslash}X
@{\hspace{4pt}}
}
\toprule
\textbf{Condition} & \textbf{Fixed $T=1$} & \textbf{Fixed $T=2$} &
\textbf{Fixed $T=3$} &
\cellcolor{tblHalting}\textbf{Adaptive} &
\cellcolor{tblHalting}\textbf{$\bar{T}$} \\
\midrule

In the Wild
& 3.39 & 1.65 & \textbf{1.31}
& \cellcolor{tblHalting}1.51
& \cellcolor{tblHalting}1.16 \\

ASVspoof 2019
& 1.05 & 1.11 & 1.34
& \cellcolor{tblHalting}\textbf{0.75}
& \cellcolor{tblHalting}1.07 \\

ASVspoof 2021 LA
& 4.51 & 4.02 & 3.99
& \cellcolor{tblHalting}\textbf{3.77}
& \cellcolor{tblHalting}1.18 \\

ASVspoof 2021 DF
& 3.13 & 3.10 & \textbf{2.96}
& \cellcolor{tblHalting}3.44
& \cellcolor{tblHalting}1.07 \\

ASVspoof 5
& 11.88 & 11.56 & \textbf{11.39}
& \cellcolor{tblHalting}11.84
& \cellcolor{tblHalting}1.19 \\

Fake or Real
& 5.60 & \textbf{0.44} & \textbf{0.44}
& \cellcolor{tblHalting}\textbf{0.44}
& \cellcolor{tblHalting}1.20 \\

CodecFake
& 1.33 & \textbf{0.18} & 0.20
& \cellcolor{tblHalting}1.11
& \cellcolor{tblHalting}1.12 \\

ADD 2022 T1
& 14.01 & 11.70 & \textbf{11.54}
& \cellcolor{tblHalting}12.43
& \cellcolor{tblHalting}1.59 \\

ADD 2022 T3
& 1.93 & 0.95 & 0.81
& \cellcolor{tblHalting}\textbf{0.74}
& \cellcolor{tblHalting}1.08 \\

ADD 2023 R1
& 1.93 & \textbf{1.81} & \textbf{1.81}
& \cellcolor{tblHalting}\textbf{1.81}
& \cellcolor{tblHalting}1.06 \\

ADD 2023 R2
& 4.58 & \textbf{4.53} & \textbf{4.53}
& \cellcolor{tblHalting}\textbf{4.53}
& \cellcolor{tblHalting}1.12 \\

DFADD
& \textbf{0.00} & \textbf{0.00} & \textbf{0.00}
& \cellcolor{tblHalting}\textbf{0.00}
& \cellcolor{tblHalting}1.00 \\

LibriSeVoc
& \textbf{0.00} & \textbf{0.00} & \textbf{0.00}
& \cellcolor{tblHalting}\textbf{0.00}
& \cellcolor{tblHalting}1.27 \\

SONAR
& 1.21 & \textbf{0.00} & 0.60
& \cellcolor{tblHalting}0.60
& \cellcolor{tblHalting}1.07 \\

\midrule

\textbf{Pooled/Macro}
& 5.32/3.90
& 4.25/2.93
& \textbf{4.19/2.92}
& \cellcolor{tblHalting}4.38/3.07
& \cellcolor{tblHalting}1.15 \\

\bottomrule
\end{tabularx}
\endgroup
\end{table*}

\begin{table*}[!t]
\centering
\caption{Complete 12-layer condition-level results. EERs are in percent.}
\label{tab:scaling_12_full}
\begingroup
\small
\setlength{\tabcolsep}{4pt}
\renewcommand{\arraystretch}{1.08}

\begin{tabularx}{\textwidth}{
@{\hspace{4pt}}
l
>{\centering\arraybackslash}X
>{\centering\arraybackslash}X
>{\centering\arraybackslash}X
>{\centering\arraybackslash}X
>{\centering\arraybackslash}X
@{\hspace{4pt}}
}
\toprule
\textbf{Condition} & \textbf{Fixed $T=1$} & \textbf{Fixed $T=2$} &
\textbf{Fixed $T=3$} &
\cellcolor{tblHalting}\textbf{Adaptive} &
\cellcolor{tblHalting}\textbf{$\bar{T}$} \\
\midrule

In the Wild
& 3.58 & 1.51 & 1.51
& \cellcolor{tblHalting}\textbf{1.19}
& \cellcolor{tblHalting}1.23 \\

ASVspoof 2019
& 1.59 & 0.82 & \textbf{0.59}
& \cellcolor{tblHalting}0.82
& \cellcolor{tblHalting}1.13 \\

ASVspoof 2021 LA
& 5.42 & 3.12 & \textbf{2.93}
& \cellcolor{tblHalting}3.06
& \cellcolor{tblHalting}1.36 \\

ASVspoof 2021 DF
& 3.22 & \textbf{2.85} & 2.96
& \cellcolor{tblHalting}3.05
& \cellcolor{tblHalting}1.10 \\

ASVspoof 5
& 10.90 & 10.10 & \textbf{10.06}
& \cellcolor{tblHalting}10.25
& \cellcolor{tblHalting}1.25 \\

Fake or Real
& 3.02 & \textbf{0.44} & 0.87
& \cellcolor{tblHalting}1.73
& \cellcolor{tblHalting}1.32 \\

CodecFake
& 1.74 & \textbf{0.24} & 0.26
& \cellcolor{tblHalting}1.41
& \cellcolor{tblHalting}1.16 \\

ADD 2022 T1
& 14.51 & \textbf{11.24} & 11.42
& \cellcolor{tblHalting}12.71
& \cellcolor{tblHalting}1.68 \\

ADD 2022 T3
& 2.22 & 0.90 & 0.85
& \cellcolor{tblHalting}\textbf{0.66}
& \cellcolor{tblHalting}1.11 \\

ADD 2023 R1
& 2.07 & 1.93 & \textbf{1.86}
& \cellcolor{tblHalting}2.06
& \cellcolor{tblHalting}1.05 \\

ADD 2023 R2
& 5.54 & \textbf{4.89} & 4.91
& \cellcolor{tblHalting}5.33
& \cellcolor{tblHalting}1.12 \\

DFADD
& \textbf{0.00} & \textbf{0.00} & \textbf{0.00}
& \cellcolor{tblHalting}\textbf{0.00}
& \cellcolor{tblHalting}1.01 \\

LibriSeVoc
& \textbf{0.00} & \textbf{0.00} & 0.13
& \cellcolor{tblHalting}\textbf{0.00}
& \cellcolor{tblHalting}1.37 \\

SONAR
& 1.21 & \textbf{0.00} & \textbf{0.00}
& \cellcolor{tblHalting}1.22
& \cellcolor{tblHalting}1.18 \\

\midrule

\textbf{Pooled/Macro}
& 5.28/3.93
& 4.19/\textbf{2.72}
& \textbf{4.17}/2.74
& \cellcolor{tblHalting}4.23/3.11
& \cellcolor{tblHalting}1.21 \\

\bottomrule
\end{tabularx}
\endgroup
\end{table*}

\clearpage
\twocolumn[{%
\begin{minipage}{\textwidth}
\centering
\captionof{table}{Hidden-state and output changes under fixed $T=3$
inference. Entries are medians with interquartile ranges in
brackets. Candidate and applied $\delta$ follow Eq.~(5).
Gate denotes the channel-mean gate value computed for each
utterance, and $|\Delta\mathrm{logit}|$ is the absolute change
in classifier logit.}
\label{tab:hidden_gate_stats}

\begingroup
\small
\setlength{\tabcolsep}{3.2pt}
\renewcommand{\arraystretch}{1.14}
\resizebox{\textwidth}{!}{%
\begin{tabular}{@{\hspace{4pt}}llcccc@{\hspace{4pt}}}
\toprule
Transition & Outcome & Candidate $\delta$ &
Applied $\delta$ & Gate
& $|\Delta\mathrm{logit}|$ \\
\midrule
$T=1\rightarrow2$ & All
& 0.299 [0.263, 0.359] & 0.116 [0.090, 0.157]
& 0.191 [0.186, 0.197] & 0.023 [0.008, 0.205] \\
$T=2\rightarrow3$ & All
& 0.231 [0.200, 0.275] & 0.037 [0.032, 0.048]
& 0.161 [0.158, 0.165] & 0.008 [0.000, 0.031] \\
\midrule
\tablebaselinerow
$T=2\rightarrow3$ & Stay correct
& 0.230 [0.199, 0.270] & 0.037 [0.032, 0.046]
& 0.161 [0.158, 0.165] & 0.008 [0.000, 0.023] \\
\tablebaselinerow
$T=2\rightarrow3$ & Stay wrong
& 0.308 [0.214, 0.514] & 0.062 [0.036, 0.108]
& 0.159 [0.153, 0.163] & 0.219 [0.062, 0.470] \\
\tablebestrow
$T=2\rightarrow3$ & Wrong $\rightarrow$ correct
& 0.464 [0.342, 0.655] & 0.130 [0.087, 0.192]
& 0.162 [0.156, 0.165] & 0.945 [0.400, 1.418] \\
\tableregressrow
$T=2\rightarrow3$ & Correct $\rightarrow$ wrong
& 0.394 [0.264, 0.575] & 0.108 [0.065, 0.177]
& 0.162 [0.161, 0.164] & 0.553 [0.227, 0.914] \\
\bottomrule
\end{tabular}%
}
\endgroup
\end{minipage}
\vspace{8pt}
}]

\subsection{Cross-depth observations}

Across all three encoder depths, two-pass refinement improves both
pooled and macro EER, but the gains vary across conditions.
A third pass produces smaller, mixed changes, consistent with the
early saturation observed for the 24-layer model. The preferred
fixed depth also differs across conditions, indicating that a
single refinement depth does not uniformly maximize detection
performance.

Adaptive inference uses fewer passes on average for the 6- and
12-layer models than for the 3-layer model, while remaining close
to fixed $T=2$ in pooled EER. However, this aggregate similarity
coexists with condition-level regressions and higher macro EER
for the 6- and 12-layer models. These results highlight the need to evaluate routing by considering
condition-level benefits and harms alongside pooled detection
performance and overall computational cost.

\section{Hidden-state and gate statistics}
\label{app:hidden_gate}

\subsection{Measurement protocol}

This appendix expands the aggregate visualization in
Figure~\ref{fig:hidden_gate} into exact distributional summaries. The
analysis uses the fixed $T=3$ trajectory of every pooled evaluation
utterance, with adaptive exits disabled, so differences between passes are
not confounded by a changing sample population. We report robust medians and
interquartile ranges because prediction-changing cases are rare and a small
number of large updates could otherwise dominate mean-based summaries.

\subsection{Metric construction and outcome groups}

The hidden-state analysis runs fixed $T=3$ inference with halting disabled,
so every utterance is observed at the same depths. Candidate change is
measured before UpdateGate, while applied change is measured after the
candidate has been blended with the preceding state. Both quantities are
computed from temporally pooled states using Eq.~\ref{eq:hidden_delta} and
are normalized by the preceding-state norm, which makes their scale
comparable across utterances. Table~\ref{tab:hidden_gate_stats} reports
medians and interquartile ranges rather than means so that a small number of
large updates do not dominate the summary.

For the outcome-conditioned rows, each pass is converted to a binary
decision using its pooled-EER threshold. ``Stay correct'' and ``stay wrong''
retain the same correctness status, while the other two categories identify
corrective and regressive flips. These categories are used only to analyze
observed trajectories; labels and future predictions do not enter model
inference.

Candidate $\delta$ measures how far the shared core proposes moving from the
previous recurrent state before gating. Applied $\delta$ measures the
corresponding realized movement after the UpdateGate mixes the proposal with
the previous state. Their ratio is therefore a descriptive measure of how
much of the proposed update is admitted.
For gate statistics, we first average the channel-wise gate values
within each utterance:
\begin{equation}
\bar{g}_{i,t}=\frac{1}{H}\sum_{h=1}^{H}g_{i,t,h}.
\end{equation}
The gate column reports the median and interquartile range of
channel-mean gate values across utterances, whereas
$|\Delta\mathrm{logit}|$ summarizes the downstream change after
ExitBridge and the frozen classifier. Gate averages describe
overall gate activation, while candidate and applied changes
measure recurrent-state update magnitudes. Keeping these
quantities separate helps distinguish state-update attenuation
from the downstream logit response.

\begin{figure*}[!t]
  \centering
  \includegraphics[width=\textwidth]{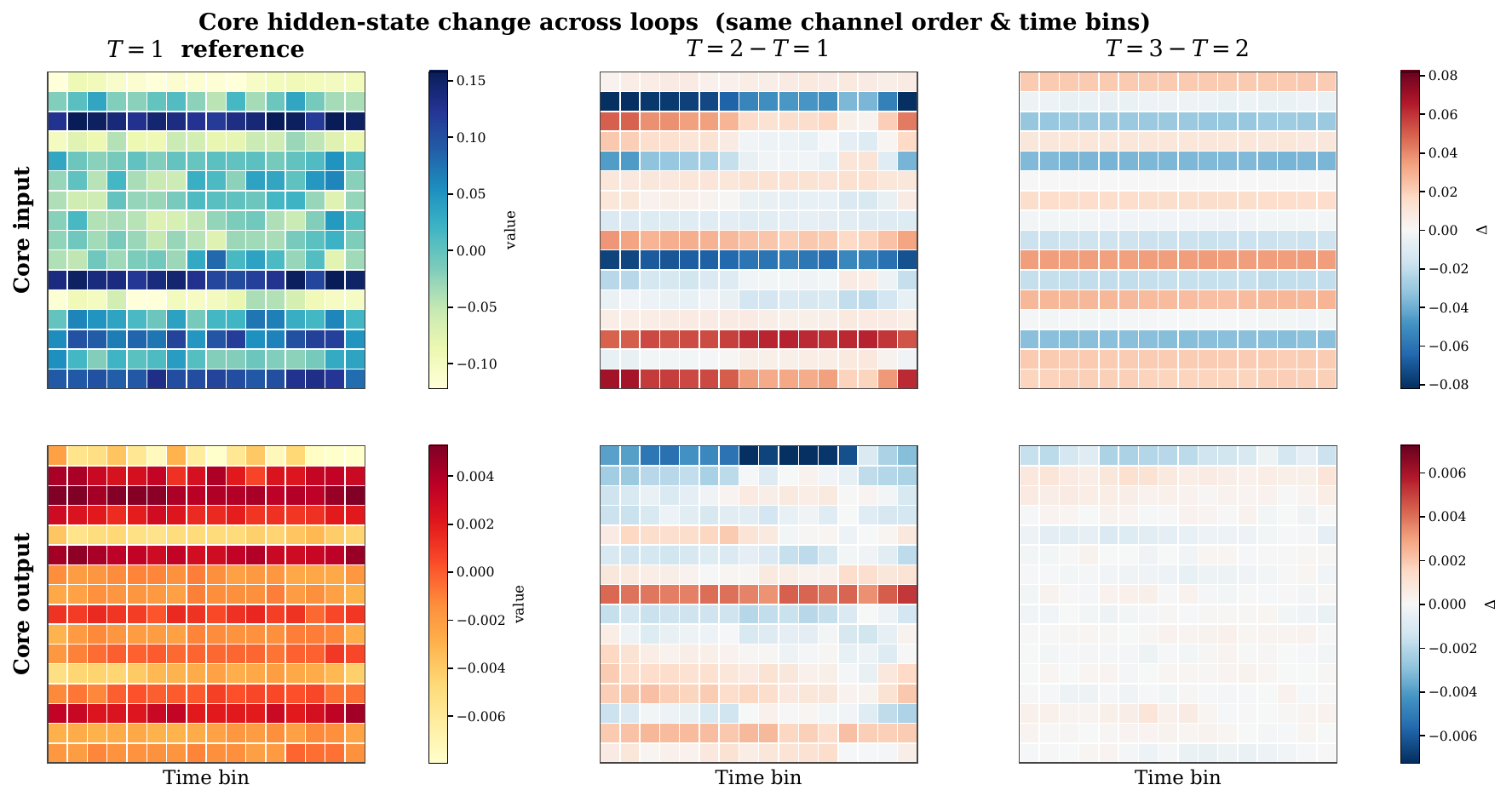}
  \caption{Core input and candidate-output evolution over 25 preselected
  diagnostic utterances. Rows use the same channel ordering and time bins;
  difference panels use row-wise symmetric color scales centered at zero.
  The subset was selected before plotting to cover stable and changing
  decisions across several test conditions, so the visualization is
  diagnostic rather than a population estimate.}
  \label{fig:core_heatmap}
\end{figure*}

\subsection{Attenuation and prediction outcomes}

The aggregate rows separate two sources of attenuation. From the second to
the third pass, the candidate update becomes smaller, but the applied update
contracts much more sharply. The gate therefore compounds a moderate change
in what the core proposes with stronger suppression of what reaches the
recurrent state. The accompanying reduction in absolute logit change
connects this representational attenuation to the small performance change
at $T=3$.

The outcome rows show that transition events are concentrated in the tail of
the update distribution. Both corrective and regressive flips have much
larger candidate and applied changes than stable-correct samples. In
contrast, their median gate values are nearly identical. This distinction is
important: the gate controls update magnitude, but whether a large update is
helpful depends on the direction of the candidate state relative to the
decision boundary. The statistics therefore explain saturation as the
combination of smaller admitted updates and increasingly mixed update
quality. They do not establish that manually increasing or decreasing the
gate would causally improve EER; such an intervention remains separate from
the observational analysis reported here.

\section{Core input--output evolution}
\label{app:core_heatmap}

\subsection{Diagnostic subset and tensor aggregation}

This analysis complements the pooled statistics with a structured view of
where successive updates occur. We use 25 utterances selected before
plotting to cover stable-correct, stable-wrong, corrective, and regressive
trajectories across several conditions. The subset is intentionally
diagnostic: it illustrates recurrent behavior but is not used to estimate
the prevalence of that behavior in the full evaluation set.

For each utterance, core-input and candidate-output tensors are aggregated
into a fixed $16\times16$ grid of channel groups and time bins. Channel
ordering and temporal partitions are shared across passes, making cell-wise
differences comparable. The reference column displays the first-pass state;
the remaining columns display successive differences with row-wise
symmetric color scales centered at zero. Keeping separate scales for core
input and output avoids letting the frontend's larger dynamic range obscure
smaller output changes.

\begin{table*}[!t]
\centering
\caption{Pooled routing diagnostics. All percentages use a fixed $0.5$
decision threshold. Recoverable recall is measured at $T=1$; harmed/continued
uses samples receiving extra computation; T1-exit correct uses samples that
exit at $T=1$.}
\label{tab:routing_quality}
\begingroup
\small
\setlength{\tabcolsep}{5pt}
\renewcommand{\arraystretch}{1.12}
\begin{tabular}{@{\hspace{4pt}}lccccc@{\hspace{4pt}}}
\toprule
Policy & $\bar{T}$ & Premature/all &
Recoverable recall & Harmed/continued
& T1-exit correct \\
\midrule
\tablebestrow
Learned halting & 1.18 & \textbf{0.30\%} & \textbf{77.10\%} & 0.89\% & \textbf{98.57\%} \\
\tablebaselinerow
Random halting & 1.75 & 0.68\% & 48.58\% & \textbf{0.34\%} & 96.16\% \\
\tablebaselinerow
Confidence@0.9 & 1.06 & 0.66\% & 48.58\% & 3.05\% & 97.53\% \\
\bottomrule
\end{tabular}
\endgroup
\end{table*}

\subsection{Quantitative reading of the heatmaps}

The visual comparison is accompanied by three scale-aware summaries. The
third-to-second update-magnitude ratio compares the norm of the later update
with that of the preceding update, so a value below one indicates
attenuation. The absolute-difference region cosine compares the nonnegative
magnitude patterns across channel--time cells; it measures whether the same
regions remain active without requiring the signed changes to point in the
same direction. Finally, the output-to-input ratio measures how much of the
change presented to the shared core appears in its candidate output. We
report medians and interquartile ranges across the 25 diagnostic utterances
for magnitude ratios, while the cosine summaries describe the aggregated
region patterns shown in the figure.

The third-to-second update-magnitude ratio is 0.578 [0.478, 0.678] at the
core input and 0.270 [0.183, 0.356] at the candidate output. Thus the third
pass begins with a smaller recurrent perturbation and the shared core
attenuates it further. The absolute-difference region cosine is 0.772 for
inputs and 0.708 for outputs. These relatively high values indicate that
the third pass largely revisits the same previously explored channel--time
regions at lower amplitude instead of redirecting computation to a
disjoint region.

We additionally measure how much input change is transferred to the core
output. The output-to-input update ratio falls from 0.151 [0.096, 0.206] at
$T=2$ to 0.079 [0.031, 0.127] at $T=3$. This drop is consistent with the
population-level reduction in candidate change and shows that attenuation
occurs inside the shared core as well as at the subsequent UpdateGate. Taken
together, the heatmap and scalar diagnostics suggest repeated refinement of
similar representational regions with progressively weaker transmission.
These observations provide a complementary view of later-pass attenuation,
linking the pooled state-change statistics to the corresponding channel--time
patterns observed in the selected diagnostic subset.

\FloatBarrier
\section{Routing-quality diagnostics}
\label{app:routing_quality}

\subsection{Definitions, thresholds, and denominators}

We align the exit decisions of learned, random, and confidence-based
policies with fixed-depth $T=1,2,3$ predictions. A first-pass error is
\emph{recoverable} if either later pass is correct. A \emph{premature exit}
stops on such an error, and \emph{continued-but-harmed} denotes a continued
sample whose final step changes a correct decision to an incorrect one.
These labels use a fixed $0.5$ classification threshold so that decisions
can be compared across passes; they are not EERs and future predictions are
used only for post-hoc analysis.

The denominators capture different failure modes. Premature/all measures
missed recoverable cases relative to the complete aligned evaluation set,
whereas recoverable recall conditions only on first-pass errors that can be
corrected at $T=2$ or $T=3$. Harmed/continued asks how often spending extra
computation ends with a correct-to-wrong transition, and T1-exit correct
measures whether computation is removed from samples that are already
solved. These quantities are complementary: a policy can exit accurately on
easy samples yet still miss the rare recoverable errors for which
further refinement matters.

The percentages in this appendix should not be compared by taking simple
complements. In particular, T1-exit correct is conditioned only on the subset
that a policy exits after the first pass, whereas baseline classification
error is measured over the complete evaluation set. The routing audit also
uses a fixed $0.5$ decision threshold to keep per-pass correctness labels
comparable. This differs from pooled EER, whose operating threshold is chosen
where false-accept and false-reject rates are equal. Thus the reported
98.57\% T1-exit correctness is a property of the selected early-exit subset,
not $100\%$ minus the 4.85\% pooled EER.

\begin{figure*}[!t]
  \centering
  \includegraphics[width=\textwidth]{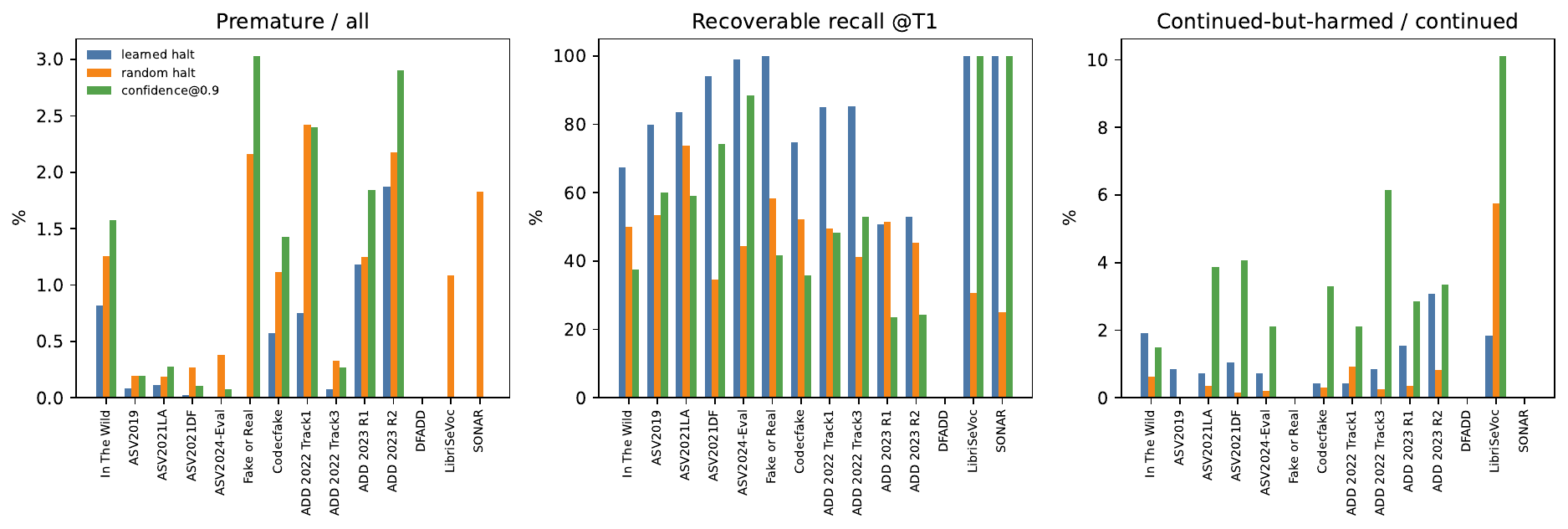}
  \caption{Per-condition routing diagnostics for learned, random, and
  confidence-based halting. Empty bars denote zero or an undefined ratio
  when a condition contains no corresponding event. All decisions use the
  fixed $0.5$ threshold described above.}
  \label{fig:routing_quality}
\end{figure*}

\subsection{Pooled routing behavior}

Table~\ref{tab:routing_quality} shows that learned halting combines accurate
early exits with stronger preservation of recoverable errors. Almost all of
its first-pass exits are already correct, while its recoverable recall is
substantially higher than either baseline. The lower premature/all rate is
especially meaningful because recoverable first-pass errors comprise only
1.25\% of the aligned set: indiscriminate early exit can easily save
computation while discarding much of this small opportunity set.

Learned halting nevertheless continues 12.73\% of samples at $T=1$, more
than the binary-recoverable fraction alone would require. This follows from
its supervision. A continuation target is positive when a future pass
reduces BCE by at least 0.01, whether or not the binary decision changes.
These BCE-improvement opportunities occur for 8.44\% of first-pass samples,
and the learned policy recalls 72.53\% of them. Moreover, among continued
samples whose decision remains unchanged, many still move farther from the
decision boundary. The policy is therefore better understood as predicting
useful refinement opportunities than as a binary error detector.

The routing is not uniformly beneficial. Continued-but-harmed events remain,
and their frequency depends on the exit rule and evaluation condition.
Figure~\ref{fig:routing_quality} shows that some domains receive substantially
more computation or have lower recoverable recall than others. A greater
average depth should therefore not be interpreted as better routing by
itself; it is useful only when the selected continuation improves the
subsequent trajectory. This variation also explains the remaining gap to the
non-deployable oracle.

\subsection{Condition-level variation}

The condition-level plots separate three questions that a single average
depth cannot answer: how often a domain receives extra computation, how many
recoverable errors are preserved, and how often continued samples are
harmed. Reading these quantities jointly avoids treating a high continuation
rate as intrinsically desirable. A policy is useful only when the extra refinement
passes are concentrated on trajectories with a favorable overall
correction--harm balance.

\FloatBarrier
\section{Dataset documentation and scope}
\label{app:scope_audit}

\begin{table*}[!t]
\centering
\caption{Training--test overlap audit for the 14 evaluation conditions.
``Unknown'' means that released metadata do not identify the generator,
not that the generator is known to be unseen.}
\label{tab:scope_audit}
\begingroup
\small
\setlength{\tabcolsep}{4pt}
\renewcommand{\arraystretch}{1.12}
\begin{tabularx}{\textwidth}{@{\hspace{4pt}}
  >{\hsize=0.65\hsize\linewidth=\hsize}X
  >{\hsize=1.40\hsize\linewidth=\hsize}X
  >{\hsize=0.95\hsize\linewidth=\hsize}X
  @{\hspace{4pt}}}
\toprule
\textbf{Condition} & \textbf{Training--test relation} &
\textbf{Evidence-backed interpretation} \\
\midrule
\rowcolors{2}{white}{tblBaseline}
ASVspoof 2019 eval & Train attacks A01--A06; eval attacks A07--A19, with
mostly familiar TTS/VC families & Generator-unseen; not uniformly
family-unseen \\
ASVspoof 5 eval & Train attacks A01--A08; eval attacks A17--A32; released
local protocol has no family field & Generator-unseen; family novelty
unknown \\
CodecFake & C1--C6 codecs occur in training; C7 (DAC) does not & Mixed:
generator-seen C1--C6 and generator-unseen C7 \\
ASVspoof 2021 LA & Dataset/channel condition is absent from training, but
attacks reuse A07--A19 from ASVspoof 2019 eval & Domain/channel-unseen;
not a new generator set beyond 2019 eval \\
ASVspoof 2021 DF & Dataset is absent from training and combines ASVspoof
and VCC sources/vocoder conditions & Domain-unseen; generator novelty is
mixed \\
In The Wild, LibriSeVoc, SONAR & Datasets are absent from adaptation
training; generator metadata are incomplete or mixed & Domain-unseen;
generator novelty not uniformly verifiable \\
DFADD & The five evaluated TTS systems also occur in adaptation training
& Generator-seen; utterance-unseen \\
Fake or Real; ADD 2022 T1/T3; ADD 2023 R1/R2 & Official held-out splits,
but released protocols do not expose generator identities &
Utterance-unseen; generator novelty unknown \\
\bottomrule
\end{tabularx}
\endgroup
\end{table*}

\subsection{Audit criteria and conservative labeling}

We audit the detector's adaptation data against each test protocol at four
levels. \emph{Utterance-unseen} means that the waveform itself is absent from
training. \emph{Generator-unseen} requires the test attack or synthesis
system to be absent from the training inventory. \emph{Family-unseen} is
stronger and requires the broader synthesis or conversion family to be new.
Finally, \emph{domain-unseen} denotes a new dataset, channel, or recording
condition, which may still contain familiar generators. These properties are
not interchangeable: a held-out dataset does not by itself establish a new
attack mechanism.

All 14 conditions are utterance-unseen because only official training splits
are used for adaptation. Stronger claims are determined by comparing the
available attack IDs, codec identities, dataset provenance, and released
metadata. When a protocol does not expose generator identity, we mark it as
unknown rather than treating missing metadata as evidence of novelty.
Table~\ref{tab:scope_audit} summarizes the strongest evidence-backed
interpretation of evaluation novelty for each group of test conditions.

The audit is relational: ``unseen'' describes a test condition only with
respect to the adaptation inventory used here. We group conditions when the
released protocols support the same conclusion and separate generator,
family, and domain novelty rather than collapsing them into a single label.
When metadata are incomplete, the table deliberately records the unsupported
dimension as unknown. This convention prevents missing metadata from being treated as evidence
for stronger claims about generalization to unseen generators.

\subsection{Data coverage and split statistics}
\label{app:artifact_documentation}

Tables~\ref{tab:training_data_statistics}
and~\ref{tab:evaluation_data_statistics} document the data sources,
language coverage, and class counts used in our experiments.
Training-source counts are obtained from the audio paths read by
the corresponding dataset loaders, before the random training--validation
split. They count loaded entries rather than globally deduplicated
waveforms, and exclude repetitions introduced by training epochs,
sampling, or online augmentation. We randomly reserve 10\% of the
loaded data for validation and use the remaining 90\% for training.
Evaluation follows the official Speech DF Arena protocols without
additional filtering or subsampling of evaluation data.

\begin{table*}[!t]
\centering
\caption{Training-source inventory before the 90/10
training--validation split. Counts refer to loaded entries.
EN and ZH denote English and Chinese, respectively.
ASVspoof counts correspond to the complete official training
partitions. Repeated source entries are included, as explained
in the text.}
\label{tab:training_data_statistics}
\begingroup
\small
\setlength{\tabcolsep}{4pt}
\renewcommand{\arraystretch}{1.12}
\begin{tabularx}{\textwidth}{
@{\hspace{4pt}}llXrrr@{\hspace{4pt}}
}
\toprule
\textbf{Dataset} &
\textbf{Language} &
\textbf{Loaded subset} &
\textbf{Bona fide} &
\textbf{Spoofed} &
\textbf{Total} \\
\midrule
\rowcolors{2}{white}{tblBaseline}
GigaSpeech &
EN &
All WAV files in the local M directory &
885,397 & 0 & 885,397 \\

SpeechFake &
EN, ZH &
\texttt{baseline/train\_all}, excluding NV &
75,708 & 468,632 & 544,340 \\

SpoofCeleb &
EN &
Training partition &
230,948 & 2,309,473 & 2,540,421 \\

DFADD &
EN &
Training lists of five synthesis systems &
214,825 & 157,500 & 372,325 \\

CN-Celeb &
ZH &
CN-Celeb2 file list &
524,787 & 0 & 524,787 \\

MLAAD &
Multilingual &
v8 synthetic speech and M-AILABS bona fide speech &
493,658 & 231,957 & 725,615 \\

Fake or Real &
EN &
\texttt{for-norm/training/fake} only &
0 & 26,927 & 26,927 \\

AISHELL-3 &
ZH &
Official training file list &
63,262 & 0 & 63,262 \\

ASVspoof 2019 LA &
EN &
Official training partition &
2,580 & 22,800 & 25,380 \\

ASVspoof 5 &
EN &
Official training partition &
18,797 & 163,560 & 182,357 \\

LibriSpeech &
EN &
\texttt{train-clean-100}, \texttt{train-clean-360},
and \texttt{train-other-500} &
281,241 & 0 & 281,241 \\

CodecFake &
EN, ZH &
CodecFake\_UCAS training partition &
105,821 & 634,926 & 740,747 \\

ADD 2023 &
ZH &
Track 1.2 and Track 3 training partitions &
6,212 & 43,269 & 49,481 \\

ADD 2022 &
ZH &
Training partition specified by its label file &
3,012 & 24,072 & 27,084 \\
\midrule
\textbf{Total} & & &
\textbf{2,906,248} &
\textbf{4,083,116} &
\textbf{6,989,364} \\
\bottomrule
\end{tabularx}
\endgroup
\end{table*}

\paragraph{Language and domain coverage.}
The inventory covers read speech, audiobooks, podcasts, online videos,
and speech from public media, together with synthesized, converted,
and codec-reconstructed audio.
The loaded SpeechFake subset contains English and Chinese:
38,400/312,111 bona fide/spoofed entries in English and
37,308/156,521 in Chinese.
CodecFake combines Chinese AISHELL-3 speech
(70,388/422,328 bona fide/spoofed entries)
and English VCTK speech (35,433/212,598 entries).

MLAAD v8 contains synthetic speech in 40 languages, identified by
\texttt{en}, \texttt{zh-cn}, \texttt{fr}, \texttt{de},
\texttt{it}, \texttt{es}, \texttt{pl}, \texttt{ru},
\texttt{nl}, \texttt{uk}, \texttt{hi}, \texttt{pt},
\texttt{tr}, \texttt{ko}, \texttt{bn}, \texttt{cs},
\texttt{hu}, \texttt{ja}, \texttt{ro}, \texttt{ar},
\texttt{fi}, \texttt{sv}, \texttt{et}, \texttt{ga},
\texttt{bg}, \texttt{mt}, \texttt{da}, \texttt{el},
\texttt{sk}, \texttt{fa}, \texttt{lt}, \texttt{hr},
\texttt{vi}, \texttt{sl}, \texttt{sw}, \texttt{th},
\texttt{tk}, \texttt{lv}, \texttt{mr}, and \texttt{ta}.
Its M-AILABS bona fide component covers German, French, Italian,
Spanish, English, Ukrainian, Polish, and Russian; English includes
both US and UK locale labels.
Thus, the language coverage of its bona fide and synthetic classes
is not identical.

\paragraph{Loading details and counting conventions.}
GigaSpeech is loaded by enumerating all WAV files in the local
M directory, without further filtering by the accompanying
metadata table. Its reported count therefore describes the
local loading inventory.
CN-Celeb uses CN-Celeb2 only.
Fake or Real contributes only the fake portion of
\texttt{for-norm/training}; its real portion is not loaded.

DFADD contains 42,965 distinct bona fide audio paths, each appearing
in the training lists of five synthesis systems.
These contribute 214,825 bona fide entries to the combined loading
inventory. This repeated inclusion is reflected in
Table~\ref{tab:training_data_statistics}; the count should not be
interpreted as the number of unique recordings.

ADD 2023 combines Track 1.2 training data
(3,012 bona fide and 24,072 spoofed entries)
with Track 3 training data
(3,200 bona fide and 19,197 spoofed entries).
For Track 3, label 6 is treated as bona fide and labels
0--5 and 7 as spoofed.
ADD 2022 uses only the training entries specified by its
training label file.

\begin{table*}[!t]
\centering
\caption{Evaluation statistics under the official Speech DF Arena
protocols. Counts refer to protocol entries rather than the full
sizes of the original dataset releases.
EN and ZH denote English and Chinese, respectively.
The SONAR language annotation is based on identifiable source
corpora and file provenance, rather than per-utterance language
annotations.}
\label{tab:evaluation_data_statistics}
\begingroup
\small
\setlength{\tabcolsep}{4pt}
\renewcommand{\arraystretch}{1.12}
\begin{tabular*}{\textwidth}{
@{\hspace{4pt}\extracolsep{\fill}}llrrr@{\hspace{4pt}}
}
\toprule
\textbf{Condition} &
\textbf{Language} &
\textbf{Bona fide} &
\textbf{Spoofed} &
\textbf{Total} \\
\midrule
\rowcolors{2}{white}{tblBaseline}
In the Wild     & EN     & 19,963  & 11,816  & 31,779 \\
ASVspoof 2019 LA & EN    & 7,355   & 63,882  & 71,237 \\
ASVspoof 2021 LA & EN    & 14,816  & 133,360 & 148,176 \\
ASVspoof 2021 DF & EN    & 14,869  & 519,059 & 533,928 \\
ASVspoof 5      & EN     & 138,688 & 542,086 & 680,774 \\
Fake or Real    & EN     & 2,264   & 2,264   & 4,528 \\
CodecFake       & EN, ZH & 92,596  & 211,645 & 304,241 \\
ADD 2022 T1     & ZH     & 31,334  & 77,865  & 109,199 \\
ADD 2022 T3     & ZH     & 20,776  & 106,085 & 126,861 \\
ADD 2023 R1     & ZH     & 80,000  & 31,976  & 111,976 \\
ADD 2023 R2     & ZH     & 87,500  & 30,977  & 118,477 \\
DFADD           & EN     & 755     & 3,000   & 3,755 \\
LibriSeVoc      & EN     & 13,201  & 79,206  & 92,407 \\
SONAR           & EN     & 2,274   & 1,674   & 3,948 \\
\midrule
\textbf{Total} & &
\textbf{526,391} &
\textbf{1,814,895} &
\textbf{2,341,286} \\
\bottomrule
\end{tabular*}
\endgroup
\end{table*}

\paragraph{Evaluation coverage and metadata limitations.}
The 14 evaluation conditions cover speech synthesis, voice conversion,
vocoder and codec artifacts, and changes in recording or transmission
conditions. Their training--test relations are separately documented
in Table~\ref{tab:scope_audit}.

The archived SONAR subset includes AudioGen, FlashSpeech,
NaturalSpeech3, OpenAI, PromptTTS2, VALLE, VoiceBox, and xTTS,
without SeedTTS. Identifiable source files point to English
corpora, including LibriSpeech, VCTK, Fisher, and LibriTTS.
The released directories do not provide per-utterance language
annotations.

The combined inventory does not provide harmonized demographic
annotations or a systematic annotation of linguistic phenomena.
We therefore report source domains and language coverage without
inferring demographic attributes from audio.
Our evaluation measures aggregate detection performance across test
conditions rather than demographic subgroup performance.

\subsection{Condition-level evidence and claim scope}

The evaluation conditions cover complementary forms of generalization.
ASVspoof 2019 and ASVspoof 5 test generator-level novelty through evaluation
attack IDs absent from their corresponding training splits, with some
attacks sharing familiar TTS and VC families. CodecFake combines codecs
represented in training (C1--C6) with the held-out C7 (DAC). ASVspoof
2021 LA and DF extend the evaluation to different channels, datasets,
and vocoder conditions. DFADD assesses generalization to held-out
utterances produced by TTS systems represented in adaptation training.

Together, these conditions evaluate detection across new generators,
domain shifts, and utterance variation. Table~\ref{tab:scope_audit}
documents this coverage and provides context for the results in
Table~\ref{tab:main}. The pooled and macro EER reductions summarize
the benefits of \method{} across this diverse evaluation suite, while
condition-level results reveal where refinement helps and where its
benefits are limited. Reading these results alongside the overlap audit
connects the observed performance to the evaluation setting of each
condition and gives a more detailed picture of generalization.

\section{Ethical Considerations}
\label{app:ethical}

\paragraph{Data and human participants.}
This study uses existing public datasets and involves no new recruitment
of human participants or human annotation.

\paragraph{Potential risks.}
False positives may undermine trust in genuine recordings, while false
negatives may allow manipulated audio to evade detection.
Detector outputs should therefore support human assessment rather than
serve as conclusive evidence of authenticity.
Improved detectors may also be misused to guide the development of
harder-to-detect deepfakes.

\paragraph{AI assistance.}
AI assistance was used for language polishing.
All algorithm design and experiments were conducted by the human authors,
who reviewed the final text and take responsibility for the manuscript.

\end{document}